\documentclass[%
 reprint,
 amsmath,amssymb,
]{revtex4-2}

\usepackage{graphicx}
\usepackage{dcolumn}
\usepackage{bm}
\usepackage{amsfonts}
\usepackage{subfigure}
\usepackage{hyperref}

\begin{document}

\preprint{APS/123-QED}

\title{Proposal for Bell test in cavity optomagnonics}

\author{Hong Xie$^{1*}$}
\author{Zhi-Gao Shi$^{1}$}
\author{Le-Wei He$^{1}$}
\author{Xiang Chen$^{2,3,4}$}
\author{Chang-Geng Liao$^{5}$}
\author{Xiu-Min Lin$^{2,3,4 \dag}$}

\affiliation{$^{1}$ Department of Mathematics and Physics, Fujian Jiangxia University, Fuzhou 350108, China}
\affiliation{$^{2}$ Fujian Provincial Key Laboratory of Quantum Manipulation and New Energy Materials, College of Physics and Energy, Fujian Normal University, Fuzhou 350117, China}
\affiliation{$^{3}$  Fujian Provincial Engineering Technology Research Center of Solar Energy Conversion and Energy Storage, Fuzhou,350117, China.}
\affiliation{$^{4}$  Fujian Provincial Collaborative Innovation Center for Advanced High-Field Superconducting Materials and Engineering, Fuzhou 350117, China}
\affiliation{$^{5}$  School of Information and Electronic Engineering (Sussex Artificial Intelligence Institute), Zhejiang Gongshang University, Hangzhou, 310018, China}

\begin{abstract}
We present a proposal to test Bell inequality in the emerging field of cavity optomagnonics, where a sphere of ferromagnetic crystal supports two optical whispering gallery modes and one magnon mode.  The two optical modes are driven by two laser pulses, respectively. Entanglement between the magnon mode  and  one of the two optical modes will be generated by the first pulse,  and the state of the magnon mode is subsequently mapped into another optical mode via the second pulse. Hence correlated photon-photon pairs are created out of the cavity. A Bell test can be implemented by using these pairs, which enables us to test the local hidden-variable models at macroscopic scales. Our results show that a significant violation of Bell inequality can be obtained in the weak-coupling regime. The violation of  Bell inequality not only verifies the entanglement between magnons and photons, but also implies that cavity optomagnonics is a promising platform for quantum information processing tasks.

\end{abstract}
\maketitle



\section{introduction}
Hybrid system enables the combination of distinct physical systems with complementary characteristics, which has played an important role in the development of quantum information \cite{duan2001long,kimble2008quantum,wehner2018quantum} and quantum sensing \cite{RevModPhys.89.035002}. In recent years, a new hybrid system based on the collective magnetic excitations in magnetic materials has emerged as a platform for novel quantum  technologies \cite{lachance2019hybrid, rameshti2021cavity}. The  quanta of the collective magnetic excitations, called magnons, have great tunability and low damping rate which make it an ideal information carrier. The magnons can interact coherently with microwave photons via magnetic dipole interaction \cite{PhysRevLett.104.077202}, and the strong coupling between magnons and microwave photons has been demonstrated experimentally with magnetic insulator yttrium iron garnet (YIG) sphere \cite{PhysRevLett.111.127003,tabuchi2014hybridizing,zhang2014strongly,PhysRevApplied.2.054002,PhysRevLett.120.057202} and stripe \cite{PhysRevLett.123.107702,PhysRevLett.123.107701}. This coupling not only allows us to study magnon-photon entanglement \cite{PhysRevLett.124.053602,PhysRevB.101.014419}, but also makes it possible to engineer an effective interaction between magnons and superconducting qubit \cite{tabuchi2015coherent,lachance2017resolving}. Benefiting from the large spin density of YIG crystal, magnons in YIG sphere can also couple with phonons through magnetostrictive interaction  \cite{zhang2016cavity,PhysRevLett.121.203601}. More recently, an exciting field named cavity optomagnonics appeared, in which a YIG sphere supports both the whispering gallery modes (WGMs) for optical photons and magnetostatic modes for magnons \cite{kusminskiy2019cavity}.
	                          
Different from the resonance interaction between microwave photons and magnons, the optomagnonic coupling between optical photons and magnons is parametrical. This is because the frequency of optical photons is in range of hundred THz, while the magnons are in GHz range. Indeed, the optomagnonic coupling  originates from magneto-optical effects, which have been used to study magnon-based microwave-optical information interconversion \cite{hisatomi2016bidirectional}. By cavity-enhanced the magnon-photon coupling in cavity optomagnonical systems, several experiments have demonstrated magnon-induced Brillouin light scattering of the optical WGMs  \cite{osada2016cavity,zhang2016optomagnonic,haigh2016triple,PhysRevLett.120.133602,PhysRevLett.123.207401}. These experiments work in the weak coupling regime, where the intrinsic optomagnonic coupling strength is much smaller than the decay rates of both optical photons and magnons. A theoretical framework for cavity optomagnonics has been established to overcome the shortcoming in this field \cite{liu2016optomagnonics,kusminskiy2016coupled,sharma2017light,graf2018cavity,osada2018orbital,haigh2018selection,sharma2019optimal, PhysRevA.104.023711}.  It is expected that the strong optomagnonic coupling will be achieved in the future, opening the door for the applications such as optical cooling the magnons \cite{sharma2018optical}, preparation of magnon Fock state \cite{bittencourt2019magnon}, magnon-based photon blockade \cite{PhysRevA.100.043831}, and magnon cat state \cite{PhysRevB.103.L100403, PhysRevLett.127.087203}.

Although it is still a challenge to realize the quantum features of cavity optomagnonics in the weak coupling regime, it is interesting that whether the nonclassicality can be charactered without quantum assumptions. Bell test is a genuine test of nonclassicality without the need of quantum formalism \cite{RevModPhys.86.419}. In this paper, we propose a proposal to violate CHSH inequality (a Bell-type inequality) \cite{PhysRevLett.23.880} by using entanglement between optical photons and magnons in a YIG sphere, which allows one to test the local hidden-variable models at macroscopic scales. The test of CHSH inequality has been performed in various systems \cite{freedman1972experimental,aspect1981experimental,shih1988new,rarity1990experimental,kwiat1995new,weihs1998violation,rowe2001experimental,hensen2015loophole,PhysRevLett.115.250401,shalm2015strong}, including recently in a macroscopic optomechanical system \cite{vivoli2016proposal,hofer2016proposal,marinkovic2018optomechanical}. However, it would be interesting to perform a Bell test in magnetically ordered solid-state system consisting of millions of spins. 

The model of our proposal involves two nondegenerate cavity modes and one magnon mode, which has been demonstrated with YIG spheres in recent experiments \cite{osada2016cavity,zhang2016optomagnonic,haigh2016triple,PhysRevLett.120.133602,PhysRevLett.123.207401}. Starting with cavity optomagnonical system close to its ground state, we use two laser pulses to excite the two cavity modes, respectively. Firstly, the optical mode 2 is driven at resonance, then the entanglement between the magnon mode and the optical mode 1 can be obtained by means of a two-mode squeezed interaction. The magnonic state can be subsequently mapped into photonic state of the optical mode 2 by a beam-splitter interaction, which is induced by driving the optical mode 1 with the second pulse.  Therefore, the photon-photon pairs of the two optical modes are generated out of the cavity optomagnonical system. The correlation of the photon-photon pairs is measured by photon detector preceded by a displacement operation in phase space. Our results show that a significant violation of CHSH inequality can be obtained in the experimentally relevant weak-coupling regime. The violation of  CHSH inequality rules out any local and realistic explanation of the measured date without quantum assumption, and it also verifies the existence of entanglement between magnons and photons.

The paper is organized as follows. The model based on cavity optomagnonics is presented in Sec.~II. Section III provides analytical discussions of the dynamical evolution of the system and the violation of CHSH inequality in phase space.   Finally, Section IV gives the conclusion.

\section{Model}
\begin{figure}[htbp]
   \subfigure{\includegraphics[width=0.45\textwidth]{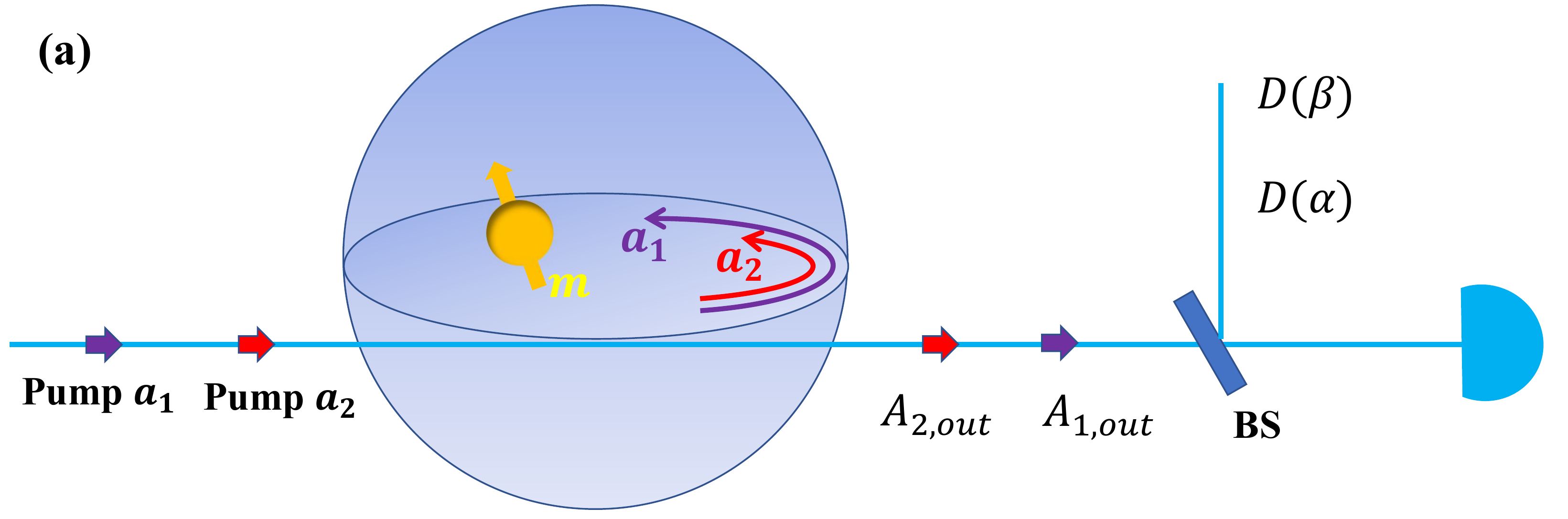}}
   \subfigure{\includegraphics[width=0.45\textwidth]{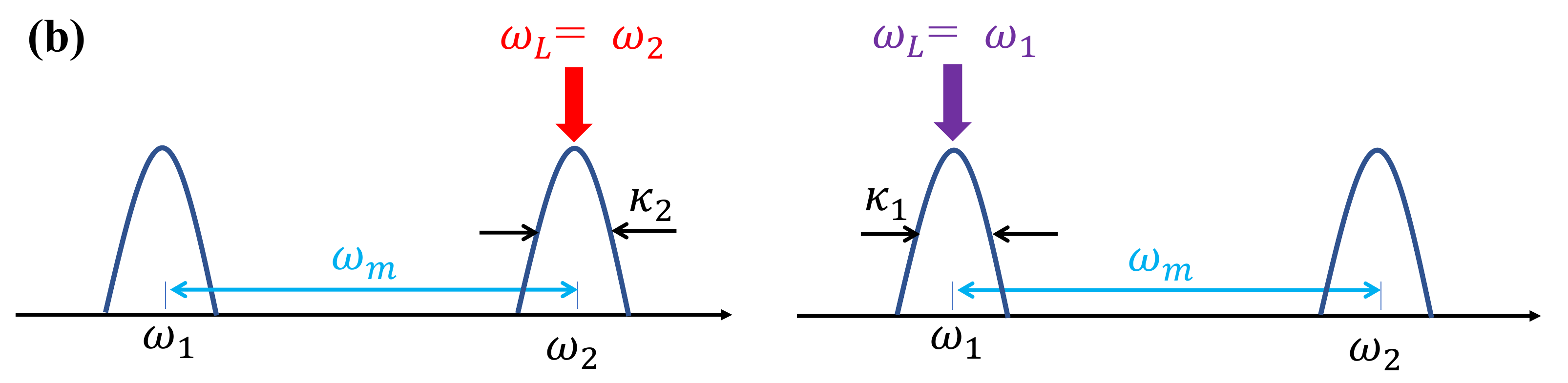}}
\caption{(Color online) (a) Schematic of cavity optomagnonical system for Bell test proposal. The first pumping pulse drives the cavity mode 2 at resonance to create entanglement between the cavity mode 1 and the magnon mode $m$. The second pulse drives the cavity mode 1 at resonance to transfer the state of the magnon mode to the cavity mode 2. The photons of cavity mode 1 and mode 2 leaking out the cavity ($A_{1,\text{out}}$ and $A_{2,\text{out}}$) are measured with a photon detector preceded by a displacement, which can be realized by an input coherent state and beam splitter (BS). (b) The two nondegenerate cavity modes $a_1$ and $a_2$ are detuned by a magnon resonance frequency $\omega_m$. }
\end{figure}

Consider  a cavity optomagnonical system where YIG sphere supports both the WGMs for optical photons and magnetostatic modes for magnons. The optomagnonic Hamiltonian is given by
\begin{eqnarray}
H= H_0 + H_{\text{int}} + H_{\text{dr}},
\end{eqnarray}
where  
\begin{eqnarray}
H_0= \omega_1 {a_1}^{\dag}a_1+\omega_2 {a_2}^{\dag}a_2+\omega_m m^{\dag}m
\end{eqnarray}
is the free evolution part of the system with bosonic operators $a_i (i=1,2)$ and $m$, $\omega_i $ and $\omega_m$ are the frequencies of the cavity modes and the magnon mode, respectively. The Hamiltonian $H_{\text{int}}$ describes the  interaction between two nondegenerate cavity modes and the magnon mode, which can be written as 
\begin{eqnarray}
H_{\text{int}}=g({a_1}{a_2}^{\dag}m+{a_1}^{\dag}a_2m^{\dag}).
\end{eqnarray}
Note that such an interaction in the optomagnonical system is subject to selection rules of angular momentum and the energy conservation requirement with $\omega_2-\omega_1=\omega_m$ \cite{haigh2016triple,sharma2017light,osada2018orbital}. It means that the creation (annihilation) of a photon in the optical mode 2 is accompanied by the annihilation (creation) of a photon in the optical mode 1 and a magnon. This optomagnonic coupling has been demonstrated experimentally in a YIG sphere, where the transverse-electric (TE) modes and the transverse-magnetic (TM) modes of the cavity interact with the magnetostatic modes  \cite{osada2016cavity,zhang2016optomagnonic,haigh2016triple,PhysRevLett.120.133602,PhysRevLett.123.207401} . When the cavity is pumped with an external field, the driving Hamiltonian reads   
\begin{eqnarray}
H_{\text{dr}}=\epsilon_i({a_i}e^{i\omega_{L}t}+{a_i}^{\dag}e^{-i\omega_{L}t}),
\end{eqnarray}
where $i=1$ or 2, $\epsilon_i$ and $\omega_{L}$ are the driving amplitudes and the driving frequencies, respectively. In the rotating frame of the driving field, the full Hamiltonian of the system becomes
\begin{eqnarray}
H= &&\Delta_1 {a_1}^{\dag}a_1+\Delta_2 {a_2}^{\dag}a_2+\omega_m m^{\dag}m \nonumber\\
&&+ g({a_1}{a_2}^{\dag}m+{a_1}^{\dag}a_2m^{\dag}) + \epsilon_i({a_i}+{a_i}^{\dag})
\end{eqnarray}
with $\Delta_i=\omega_i-\omega_{L}$.

In the case of that the cavity mode 2 is pumped and the cavity mode 1 is undriven, following the standard linearization procedure, we split both the cavity modes and the magnon mode into an average amplitude and a fluctuation term, i.e., $a_i\rightarrow \alpha_i + a_i$ and $m\rightarrow \beta + m$. The average amplitude can be obtained as $\alpha_2=\epsilon_2/(i\kappa_{2,\text{ex}}/2-\Delta_2)$ and $\beta=0$, where $\kappa_{i,\text{ex}}$ denotes the loss rate of the $i$th cavity mode associated with the external coupling. Since only the mode 2 is pumped we have $\alpha_1=0$. Note that the coherent amplitude $\alpha_i$ can be chosen real by a suitable choice of the phase of the pumping field. The linearized Hamiltonian of the system in the interaction picture can be gained
\begin{eqnarray}
H_{\text{int1}}= G_1({a_1}m e^{i(\Delta_1+\omega_m)t}
+{a_1}^{\dag}m^{\dag}e^{-i(\Delta_1+\omega_m)t}),
\end{eqnarray}
where $G_1=g\alpha_2$, and the small nonlinear term has been neglected. When the cavity mode 2 is driven at resonance $\omega_{L}=\omega_2$ such that $\Delta_1=-\omega_m$, Eq.~(6) becomes
\begin{eqnarray}
H_{\text{int1}}= G_1({a_1}m + {a_1}^{\dag}m^{\dag}),
\end{eqnarray}
which represents a two-mode squeezing interaction. Assumed that both the optical modes and the magnon mode are initially prepared in their ground state, the entanglement between the cavity mode 1 and the magnon mode can be generated after the pulse duration. In the weak coupling regime, i.e., the effective optomagnonic coupling strength $G_1$ is much smaller than the decay rate $\kappa_1$ of the cavity, the entangled photons leak out the cavity faster than they are generated, therefore the magnon mode turns to entangle with a traveling optical pulse. Here, we will show the entangled states of the traveling-wave optical fields and the magnon mode can be utilized to test the Bell inequality.

To perform a measurement on the magnon mode, the magnon state should be transferred to the optical field. We now consider that only the cavity mode 1 is pumped, similar to the process of pumping the mode 2 discussed above, we can obtain the interaction Hamiltonian
\begin{eqnarray}
H_{\text{int2}}= G_2({a_2}m^{\dag} e^{-i(\Delta_2-\omega_m)t}
+{a_2}^{\dag}me^{i(\Delta_2-\omega_m)t})
\end{eqnarray}
with $G_2=g\alpha_1$, and the average amplitude of the mode 1 is given by $\alpha_1=\epsilon_1/(i\kappa_{1,\text{ex}}/2-\Delta_1)$. When the mode 1 is driven at resonance $\omega_L=\omega_1$ so that $\Delta_2=\omega_m$, the Hamiltonian becomes
\begin{eqnarray}
H_{\text{int2}}= G_2({a_2}m^{\dag} + {a_2}^{\dag}m).
\end{eqnarray}
The Hamiltonian is often referred to as a beam-splitter interaction, which is relevant for the state transfer between the cavity mode 2 and the magnon mode.

Note that the coupling between the cavity mode 1 (2) and the magnon mode is achieved by pumping the other cavity mode 2 (1). The proposal for Bell test in cavity optomagnonics is depicted in Fig.~1. It can be summarized as three steps. (i) The entanglement between the cavity mode 1 and the magnon mode is generated by resonantly pumping the mode 2. (ii) The quantum state of the magnon mode is subsequently mapped into the cavity mode 2 by resonantly driving the mode 1, therefore the modes 1 and 2 are entangled.  (iii) The measurement for the correlation between the two traveling optical pulses of the modes 1 and 2 is performed.  The measurement setting consists of a single-photon detector preceded by a displacement operation $D(\alpha)$, which can be implemented by an input coherent state and an unbalanced beam splitter  \cite{paris1996displacement}. Such measuring apparatuse has been used for Bell tests in optical experiments \cite{kuzmich2000violation}.

\section{Bell test in cavity optomagnonics}
\subsection{Generation of optomagnonical entanglement}
In this section, we study the dynamics evolution of the system and the generation of optomagnonical entanglement for Bell test. When the cavity mode 2 is driven at resonance, the quantum Langevein equation with the Hamiltonian $H_\text{{int1}}$ are
\begin{subequations}
\begin{equation}
\frac{da_1}{dt}=-iG_1m^{\dag}-\frac{\kappa_{1}}{2}a_1+\sqrt{\kappa_{1,\text{ex}}}a_{1,\text{in}}+\sqrt{\kappa_{1,\text{i}}}a_{1,\text{th}},
\end{equation}
\begin{eqnarray}
\frac{dm}{dt}=-iG_1{a_1}^{\dag}-\frac{\gamma}{2}m+\sqrt{\gamma}m_{\text{in}},
\end{eqnarray}
\end{subequations}
where $\kappa_1=\kappa_{1,\text{ex}}+\kappa_{1,\text{i}}$ is the total linewidth of the cavity mode 1 in terms of the external coupling $\kappa_{1,\text{ex}}$ and material optical loss $\kappa_{1,\text{i}}$,  $\gamma$ denotes decay rate of the magnon mode. $a_{1,\text{in}}$ is the vacuum input noise for cavity,  $a_{1,\text{th}}$  denotes the thermal noise introduced by material and $m_{\text{in}}$ represents the stochastic magnetic field. The noise correlators associated with the memoryless Markov-like fluctuations are $\langle X^{\dag}(t)X(t')\rangle=\bar{n}_\text{th}\delta(t-t')$ and $\langle X(t)X^{\dag}(t')\rangle=(\bar{n}_\text{th}+1)\delta(t-t')$, where $X\in \{a_{1,\text{in}}, a_{1,\text{th}}, m_{\text{in}}\}$. The optical field has zero thermal occupation ($\bar{n}_\text{th} \approx 0$) even at room temperature, but this is not the case for magnons.

For simplicity, we consider the total loss of the cavity is dominated by the external coupling $\kappa_1\approx \kappa_{1,\text{ex}} \gg \kappa_{1,\text{i}}$, i.e., the cavity is over-coupled. Assuming that the magnons  is initially prepared in its ground state, then the noise introduced by material  is negligibly small. After neglecting the decay of the magnon mode, which is reasonable when the pulse duration is shorter than the magnon decoherence time $(\gamma \bar{n}_{\text{th}})^{-1}$, the quantum Langevein equations can be simplified as
\begin{subequations}
\begin{equation}
\frac{da_1}{dt} \approx -iG_1m^{\dag}-\frac{\kappa_{1}}{2}a_1+\sqrt{\kappa_{1}}a_{1,\text{in}},
\end{equation}
\begin{eqnarray}\label{mderivation}
\frac{dm}{dt} \approx -iG_1{a_1}^{\dag},
\end{eqnarray}
\end{subequations}

In the weak coupling regime $G_1\ll \kappa_1$, the cavity is damped at a rate that is much faster than the rate at which the magnon mode changes the cavity state, then the cavity will reach a quasistationary state, which quickly adjust to changes in magnon mode. Therefore, the mode $a_1$ can be adiabatically eliminated by setting $da_1/dt=0$, and we have
\begin{eqnarray}\label{a1aida}
a_1\approx (-i2G_1/\kappa_1)m^{\dag}+(2/\sqrt{\kappa_1})a_{1,\text{in}}.
\end{eqnarray}
The white noise $a_{1,\text{in}}$ has infinite bandwidth, it is not strictly possible to slave $a_1$ that only responds to a finite bandwidth $\kappa_1$, to the vacuum fluctuation. However, as far as the magnon mode $m$ is concerned, vacuum fluctuation restricted to a finite bandwidth can be effectively treated as white noise, hence there is no harm that $a_1$ is slaved to the vacuum fluctuation \cite{PhysRevA.49.4110}. Although the noise $a_{1,\text{in}}$ restricted to finite bandwidth in the adiabatic elimination that leads to an incorrect communication relations of $a_1$, it is known that only the frequency near resonance with the system are important, the high frequency part of noise which far from resonant are contribute a little when tracing over the bath. Hence the adiabatic elimination of mode $a_1$ can simplify the calculation while maintaining the physical reliability.

Substituting the expression of $a_1$ into the input-output relation $a_{1,\text{out}}=-a_{1,\text{in}}+\sqrt{\kappa_1}a_1$ and Eq.~(\ref{mderivation}), we get
\begin{subequations}
\begin{eqnarray}\label{aout}
a_{1,\text{out}}=a_{1,\text{in}}-i\sqrt{2\tilde{G}_1}m^{\dag},
\end{eqnarray}
\begin{eqnarray}\label{dm}
\frac{dm}{dt}=\tilde{G}_1m-i\sqrt{2\tilde{G}_1}a^{\dag}_{1,\text{in}},
\end{eqnarray}
\end{subequations}
where $\tilde{G}_1=2{G_1}^2/\kappa_1$. To solve these equations, it is convenient to introduce the effective temporal modes \cite{hofer2011quantum,galland2014heralded}. For the cavity mode 2 is driven by a pulse with the duration $t=\tau_1$, the effective temporal modes read
\begin{subequations}
\begin{equation}
A_{1,\text{in}}(\tau_1)=\sqrt{\frac{2\tilde{G}_1}{1-e^{-2\tilde{G}_1\tau_1}}}\int_0^{\tau_1}
{dte^{-\tilde{G}_1t}a_{1,\text{in}}(t)},
\end{equation}
\begin{equation}
A_{1,\text{out}}({\tau_1})=\sqrt{\frac{2\tilde{G}_1}{e^{2\tilde{G}_1{\tau_1}}-1}}\int_0^{\tau_1}
{dte^{\tilde{G}_1t}a_{1,\text{out}}(t)}.
\end{equation}
\end{subequations}
Then the explicit solution of  Eq.~(\ref{dm})  can be expressed as
\begin{equation}\label{mout}
m({\tau_1})=e^{\tilde{G}_1{\tau_1}}m(0)-i\sqrt{e^{2\tilde{G}_1{\tau_1}}-1}A^{\dag}_{1,\text{in}}({\tau_1}).
\end{equation}
Note that Eq.~(\ref{dm}) can be rewritten as $a_{\text{1,in}}^\dag = (1/i\sqrt{2\tilde{G}_1})(\tilde{G}_1m-\frac{dm}{dt})$. Substitute it into the Eq.~(\ref{aout}), we obtain the equation of motion $\frac{dm}{dt}+\tilde{G}_1m = -i\sqrt{2\tilde{G}_1}a_{1,\text{out}}^\dag$, which relating the dynamic evolution of magnons to the output optical field. The solution of this equation is given by
$m({\tau_1})=e^{-\tilde{G}_1{\tau_1}}m(0)-i\sqrt{e^{1-2\tilde{G}_1{\tau_1}}}A^{\dag}_{1,\text{out}}({\tau_1})$. Combining it with Eq.~(\ref{mout}), we have
\begin{equation}\label{Aout}
A_{1,\text{out}}({\tau_1})=e^{\tilde{G}_1{\tau_1}}A_{1,\text{in}}({\tau_1})-i\sqrt{e^{2\tilde{G}_1{\tau_1}}-1}m^{\dag}(0).
\end{equation}

The solutions (\ref{mout}) and (\ref{Aout}) can be written as $A_{\text{1,out}}=U_1^
{\dag}({\tau_1})A_{\text{1,in}}U_1({\tau_1})$ and $m({\tau_1})=U_1^{\dag}({\tau_1})m(0)U_1({\tau_1})$, where the propagator $U_1({\tau_1})$ is extracted as  \cite{galland2014heralded}
\begin{equation}
U_1({\tau_1})=e^{-i\sqrt{p}A_{\text{1,in}}^{\dag}m^{\dag}}e^{-\tilde{G}_1\tau_1(1+A_{\text{1,in}}^{\dag}A_{\text{1,in}}+m^{\dag}m)}e^{i\sqrt{p}A_{\text{1,in}}m}   
\end{equation}
with $p={1-e^{-2\tilde{G}_1\tau_1}}$.  Asssumed that the system is initially in the vacuum state $\rho_0=\left|000\right>_{A_1A_2m}\left<000\right|$, at the end of the pumping pulse, the system evolves into 
\begin{equation}\label{rho1}
\rho_1=U_1({\tau_1})\left|000\right>_{A_1A_2m}\left<000\right|U_1^{\dag}({\tau_1}). 
\end{equation}
Note that the operators $A_{\text{1,in}}m$, $A_{\text{1,in}}^{\dag}A_{\text{1,in}}$, and $m^{\dag}m$ have zero eigenvalue for the state $\left|000\right>$, so that $\rho_1 = e^{-2\tilde{G}_1\tau_1} e^{-i\sqrt{p}A_{\text{1,in}}^{\dag}m^{\dag}} \left|000\right>_{A_1A_2m}\left<000\right|e^{i\sqrt{p}A_{\text{1,in}}m}  $. Then we have
\begin{equation}\label{rho11}
\rho_1 = (1-p) \sum_{n,n'=0}^{\infty}{(-1)^n(i\sqrt{p})^{n+n'} \left|n,0,n\right>_{A_1A_2m}\left<n',0,n'\right|}.
\end{equation}
It is clear that the optical mode 1 and the magnon mode are entangled, and the optical mode 2 stays in the vacuum state. We introduce formally $p= 1-e^{-2\tilde{G}_1\tau_1} = \tanh^2 r$ and $1-p = e^{-2\tilde{G}_1\tau_1} = \cosh^{-2} r$, where $r$ denotes the squeezing parameter. Then the state of the mode 1 and the magnon mode becomes $\left| \Psi \right>_{A_1,m}=\cosh^{-1}r\sum_{n=0}^{\infty}(-i)^n\tanh^nr \left|n,n\right>_{A_1,m}$, which is the standard form of the two-mode squeezed state.

In order to test Bell inequality by using the entanglement between the mode 1 and the magnon mode $m$, the magnonic state  should be transferred to the optical mode 2 for the measurement purpose. We now turn to consider that the cavity mode 1 is pumped by the second pumping pulse. In this case, the dynamics of the mode 2 and the magnon mode is described by Hamiltonian $H_\text{{int2}}$. The corresponding simplified quantum Langevin equations are
\begin{subequations}
\begin{equation}
\frac{da_2}{dt} \approx -iG_2m-\frac{\kappa_2}{2}a_2+\sqrt{\kappa_2}a_{2,\text{in}},
\end{equation}
\begin{equation}
\frac{dm}{dt} \approx-iG_2{a_2}.
\end{equation}
\end{subequations}
Following the same procedures discussed as for the mode 1 and the magnon mode, we can obtain $a_{2,\text{out}}=a_{2,\text{in}}-i\sqrt{2\tilde{G}_2}m$,
and $\frac{dm}{dt}=-\tilde{G}_2m-i\sqrt{2\tilde{G}_2}a_{1,\text{in}}$, where $\tilde{G}_2=2{G_2}^2/\kappa_2$. By introducing the effective temporal modes of the cavity mode 2
\begin{subequations}
\begin{equation}
A_{2,\text{in}}({\tau_2})=\sqrt{\frac{2\tilde{G}_2}{e^{2\tilde{G}_2{\tau_2}}-1}}\int_0^{\tau_2}
{dte^{\tilde{G}_2t}a_{2,\text{in}}(t)},
\end{equation}
\begin{equation}
A_{2,\text{out}}({\tau_2})=\sqrt{\frac{2\tilde{G}_2}{1-e^{-2\tilde{G}_2{\tau_2}}}}\int_0^{\tau_2}
{dte^{-\tilde{G}_2t}a_{2,\text{out}}(t)}
\end{equation}
\end{subequations}
with the duration ${\tau_2}$ of pumping pulse, we have
\begin{subequations}
\begin{equation}\label{A2out}
A_{2,\text{out}}({\tau_2})=e^{-\tilde{G}_2{\tau_2}}A_{2,\text{in}}({\tau_2})-i\sqrt{1-e^{-2\tilde{G}_2{\tau_2}}}m(0),
\end{equation}
\begin{equation}\label{m2out}
m({\tau_2})=e^{-\tilde{G}_2{\tau_2}}m(0)-i\sqrt{1-e^{-2\tilde{G}_2{\tau_2}}}A_{2,\text{in}}({\tau_2}).
\end{equation}
\end{subequations}
By rewriting the solutions (\ref{A2out}) and (\ref{m2out}) as $A_{\text{2,out}}=U_2^
{\dag}(\tau_2)A_{\text{2,in}}U_2(\tau_2)$ and $m(\tau_2)=U_2^{\dag}(\tau_2)m(0)U_2(\tau_2)$, the propagator $U_2(\tau_2)$ can be obtained as
\begin{equation}
U_2(\tau_2)=e^{-i\sqrt{T'}A_{\text{2,in}}^{\dag}m}e^{\tilde{G}_2\tau_2(A^{\dag}_{\text{2,in}}A_{\text{2,in}}-m^{\dag}m)}e^{i\sqrt{T'}A_{\text{2,in}}m^{\dag}}   
\end{equation}
with $T'=e^{2\tilde{G}_2\tau_2}T$, where $T={1-e^{-2\tilde{G}_2\tau_2}}$ denotes the conversion efficiency between the magnon mode $m$ and the mode 2. If $\tilde{G}_2\tau_2$ is sufficient large,
Eq.~(\ref{A2out}) is reduced to $A_{2,\text{out}}(\tau_2)=-im(0)$. This means that the magnon quantum state created by the first pulse can be nearly perfectly mapped onto the optical mode 2 apart from a phase shift. When the two cavity modes and the magnon mode are initially in vacuum state, by sequentially pumping the two cavity modes at resonance, the final state of the system can be described by density matrix 
\begin{equation}
\rho_2=U_2({\tau_2})U_1({\tau_1})\left|000\right>_{A_1A_2m}\left<000\right|U_1^{\dag}({\tau_1})U_2^{\dag}({\tau_2}).
\end{equation}
Recalling the Eqs.~(\ref{rho1}) and (\ref{rho11}), the density matrix can be written as $\rho_2=U_2({\tau_2})\rho_1U_2^{\dag}({\tau_2})$. Then, the density matrix is calculated as
\begin{equation}
\rho_2=(1-p) \sum_{n,n'=0}^{\infty}{(-\sqrt{pT})^{n+n'} \left|n,n,0\right>_{A_1A_2m}\left<n',n',0\right|}.
\end{equation}

By tracing out the magnon mode, we gain the density matrix of the two travelling optical pulses 
\begin{equation}\label{rhoA1A2}
\rho_{A_1A_2}=(1-p) \sum_{n=0}^{\infty}{(-\sqrt{pT})^{n+n'} \left|n,n\right>_{A_1A_2}\left<n',n'\right|}.
\end{equation}
In the case of $\tilde{G}_2\tau_2 \gg 1$, the conversion efficiency $T$ approaches to $1$, and the state $\rho_{A_1A_2}$ is close to two-mode squeezed state.

\subsection{Violation of CHSH inequality}
The type of Bell inequality relevant to our proposal is the CHSH inequality \cite{PhysRevLett.23.880}. We are interesting in the measurements that allow us to identify the vacuum state  and all nonzero photon number states, i.e., the on-off detection. When the application of coherent displacement $D(\alpha)$ is in front of the photon detector, the measurement can be described by the positive-operator-valued measure with two orthogonal projection operators  $P_{\alpha}=D(\alpha)\left|0 \right>\left<0 \right|D^{\dag}(\alpha)=\left|\alpha \right>\left<\alpha \right|$ and $Q_{\alpha}=\mathbb{I}-\left|\alpha \right>\left<\alpha \right|$. We assign the outcome $+1$ to the detection of $\left|\alpha \right>$ and $-1$ otherwise, then the observable for the system is described by $2P_{\alpha}-\mathbb{I}$. Therefore, the correlation function $E_{\alpha\beta}=\left< (2P_{\alpha}-\mathbb{I})\otimes(2P_{\beta}-\mathbb{I})\right>$ for the two optical fields is given by
 \begin{equation}
 E_{\alpha\beta}= 4P(+1+1|\alpha\beta)-2[P(+1|\alpha)+P(+1|\beta)]+1.   
\end{equation}
Here $P(+1+1|\alpha\beta)=\left<P_{\alpha} \otimes P_{\beta}\right>$ represents the joint probability to get measurement outcome $+1$ for both optical fields, $P(+1|\alpha)=\left<P_{\alpha} \otimes \mathbb{I}\right>$ and $P(+1|\beta)=\left< \mathbb{I}\otimes P_{\beta}\right>$ are the probabilities of measuring single field with outcome $+1$, respectively. The observable and correlation of such form for Bell tests were introduced in Refs.\cite{tan1991nonlocality,banaszek1999testing,hardy1994nonlocality,hessmo2004experimental}, and had been realized in experiments with optical two-mode squeezed states produced by spontaneous parametric down conversion \cite{kuzmich2000violation}. 

For the local hidden-variable model, the four correlation functions between pairs of measurements obey the CHSH inequality
\begin{equation}\label{CHSH}
S=| E_{{\alpha_1}{\beta_1}}+E_{{\alpha_1}{\beta_2}}+E_{{\alpha_2}{\beta_1}}-E_{{\alpha_2}{\beta_2}}|\leq 2.   
\end{equation}
The inequality can be violated with a proper choice of the observables measured on the quantum entanglement states, and the allowed maximal violation is $S=2\sqrt{2}$ \cite{RevModPhys.86.419}.

\begin{figure}[!htb]
\includegraphics[width=0.48\textwidth]{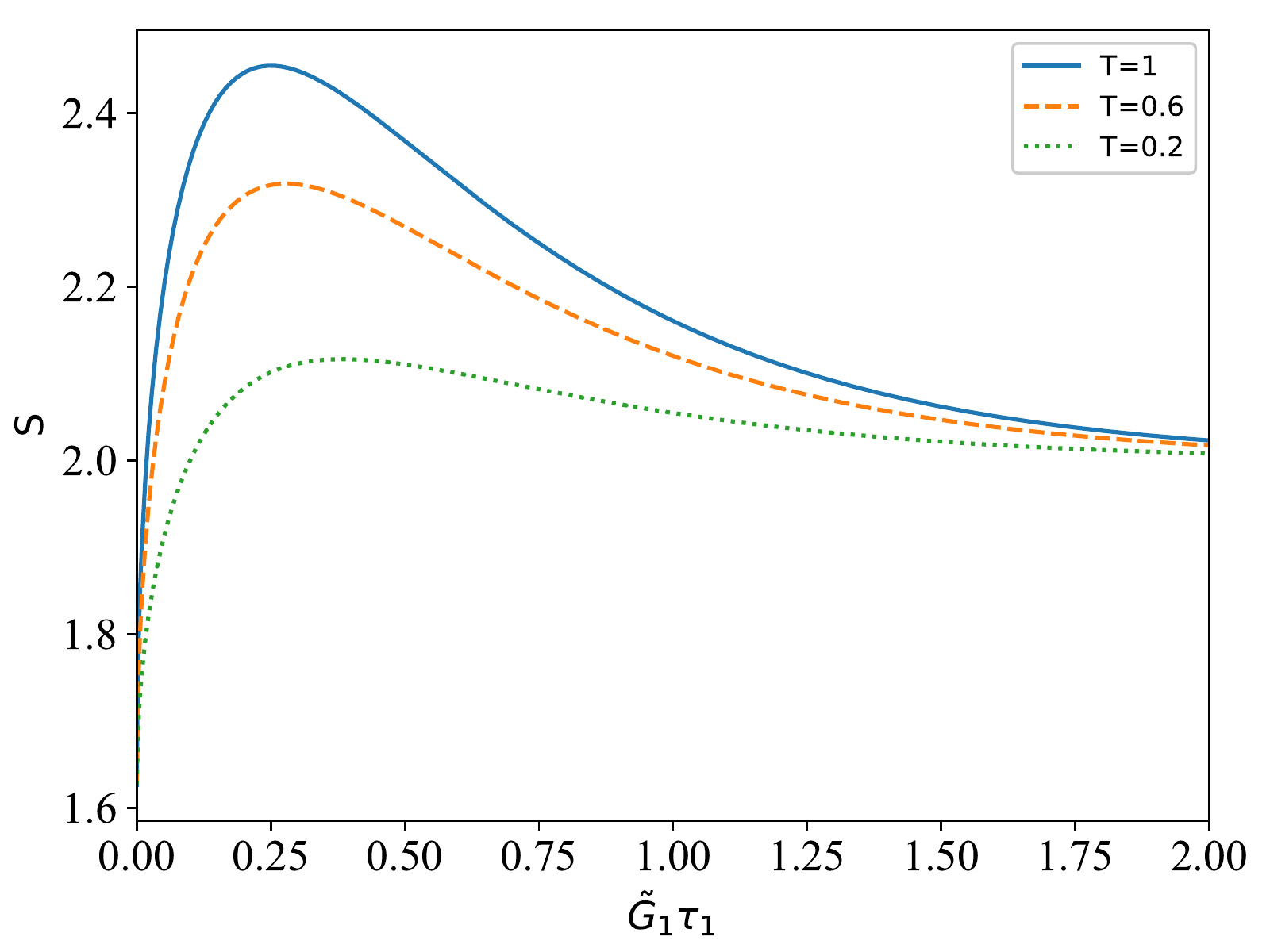}
\caption{Optimal values of $S$ as a function of $\tilde{G}_1\tau_1$ for various magnon-photon conversion efficiencies $T$.}
\end{figure}
\begin{figure}[htb]
\includegraphics[width=0.48\textwidth]{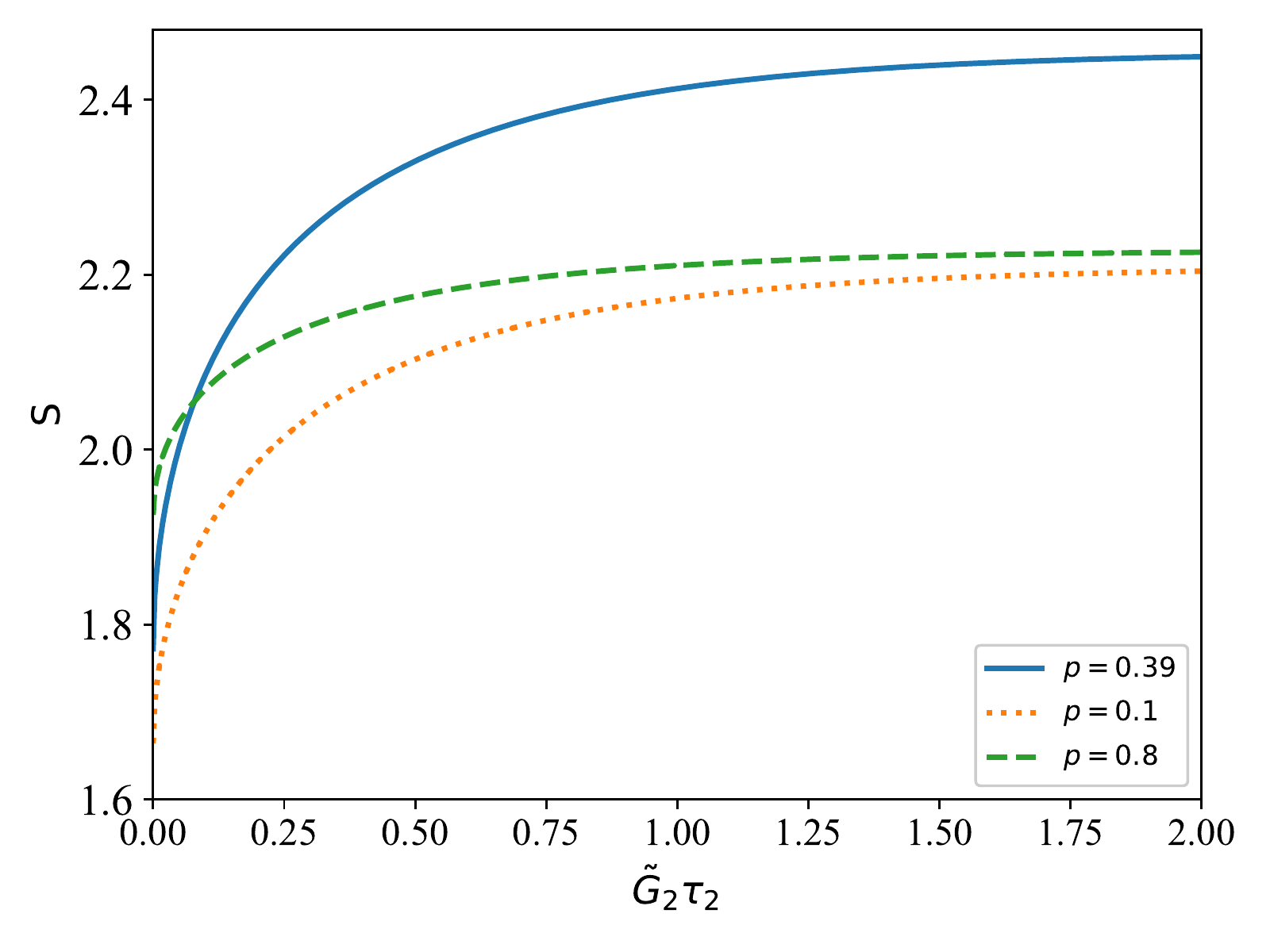}
\caption{Optimal values of $S$ as a function of $\tilde{G}_2\tau_2$ for various values of $p$.}
\end{figure}
We now proceed to discuss the correlation function $E_{\alpha\beta}$ between the optical modes 1 and 2. The joint probability of measurement outcomes $+1$ for both modes 1 and 2 can be written as $P(+1+1|\alpha\beta)=\text{Tr}(\rho_{A_1A_2}P_{\alpha}\otimes P_{\beta})$.  By using the density matrix given by Eq.~(\ref{rhoA1A2}), we calculate the joint probability as
\begin{equation}
P(+1+1|\alpha\beta)= (1-p)e^{-|\alpha|^2-|\beta|^2}e^{-\sqrt{pT}(\alpha^*\beta^*
+\alpha\beta)}.
\end{equation}
The marginals $P(+1|\alpha)=\text{Tr}({\rho}_{A_1A_2}P_{\alpha}\otimes \mathbb{I})$ and
$P(+1|\beta)=\text{Tr}({\rho}_{A_1A_2} \mathbb{I} \otimes P_{\beta})$ are also given by
\begin{equation}
P(+1|\alpha)= (1-p)e^{-(1-pT)|\alpha|^2}
\end{equation}
and
\begin{equation}
P(+1|\beta)= (1-p)e^{-(1-pT)|\beta|^2},
\end{equation}
respectively. Together with the definition of correlation function $E_{\alpha\beta}$, the quantity S can be evaluated by Eq.~(\ref{CHSH}).

We optimize the value of $S$ over the measurement settings $\alpha_{1,2}$ and $\beta_{1,2}$, and the results as a function of $\tilde{G}_1\tau_1$ for different conversion efficiencies $T$ are shown in Fig.~2. Obviously, the violation of CHSH inequality can be obtained with a proper choice of $\tilde{G}_1\tau_1$ and a high conversion efficiency $T$. The maximal violation $S\approx 2.45$ is achieved at $\tilde{G}_1\tau_1 \approx 0.25$ ($p\approx0.39$) and $T=1$. This result agrees well with previous studies \cite{lee2009testing,brask2012robust}, where the maximal violation of $S\approx 2.45$ for squeezing parameter $r \approx 0.76$ ($p=\tanh^2 r \approx 0.40$) can be obtained for an ideal two-mode squeezed state.

The quantity $S$ versus $\tilde{G}_2\tau_2$ for different values of $p$  is depicted in Fig.~3.  It is shown that a significant violation is achieved at  $p=0.39$ with relative larger $\tilde{G}_2\tau_2$.
An efficient conversion requires stronger coupling strength ${G}_2$ between the magnon mode and the cavity mode 2, which can be obtained by increasing the input pumping power encoded in $\alpha_1$.

Recalling the approximation have been made in the model: (i) weak coupling conditions, $G_{1,2} \ll \kappa$; (ii) negligible magnon dissipation which implies the pulse duration should be shorter than the magnon decoherence time $\tau_1+ \tau_2 \ll (\gamma\bar{n}_\text{th})^{-1}$. In the YIG-based cavity optomagnonical system, the cavity decay rate $\kappa\sim 1$ \text{GHz} and magnon decay rate $\gamma\sim  1$  \text{MHz} has been demonstrated in the experiment \cite{zhang2016optomagnonic,haigh2016triple}. Assuming that the coupling strengths are $G_{1}\sim 20$ \text{MHz} and $G_{2}\sim 100$ \text{MHz},  with the pulse durations $\tau_1\sim 31$ \text{ns} and $\tau_2\sim 75$ \text{ns}, the optimal value of $\tilde{G}_1\tau_1$ around $0.25$ and high conversion efficiency $T ={1-e^{-2\tilde{G}_2\tau_2}}\approx 0.95$ can be achieved. In this case, the requirement imposed by the two approximations above can be satisfied. Although the required coupling strength are still unavailable in current experiments, great efforts have been paid to enhance the optomagnonic coupling by reducing the mode volume of the optical mode in YIG disk \cite{graf2018cavity}, or by selecting magnon modes to maximize the overlap of magnon and photon modes \cite{sharma2019optimal}. It is to be expect that the coupling strength will be improved in the next generation experiments.

\subsection{Bell test in phase space}
In above discussions, we have neglected the influence on measurement for the efficiency of photon detector and the transmissivity in the beam splitter. In the following, the proposal for Bell test including the detector efficiencies is discussed in phase space. For the measurement setting in our proposal, it has been shown that the Bell inequality can be studied in the phase space based on Wigner function or Q function \cite{banaszek1999testing}.  For the on-off detection, which measures the correlation between the vacuum state  and all nonzero photon number states,  the mean value of the measurement is proportional to the Q functions with $Q(\alpha,\beta)=\frac{1}{\pi^2}P(+1+1|\alpha\beta)$ and $Q(\alpha)=\frac{1}{\pi}P(+1|\alpha)$. The  CHSH inequality can be formulated in terms of the Q functions as \cite{banaszek1999testing,li2017einstein}
\begin{eqnarray}\label{CHSHQ}
S=&&| 4\pi^2[Q({{\alpha_1},{\beta_1}})+Q({{\alpha_1},{\beta_2}})+Q({{\alpha_2},{\beta_1}})-Q({{\alpha_2},{\beta_2}})] \nonumber\\
&&-4\pi[Q(\alpha_1)+Q(\beta_1)] + 2|\leq 2.   
\end{eqnarray}

\begin{figure}[htb]
\includegraphics[width=0.48\textwidth]{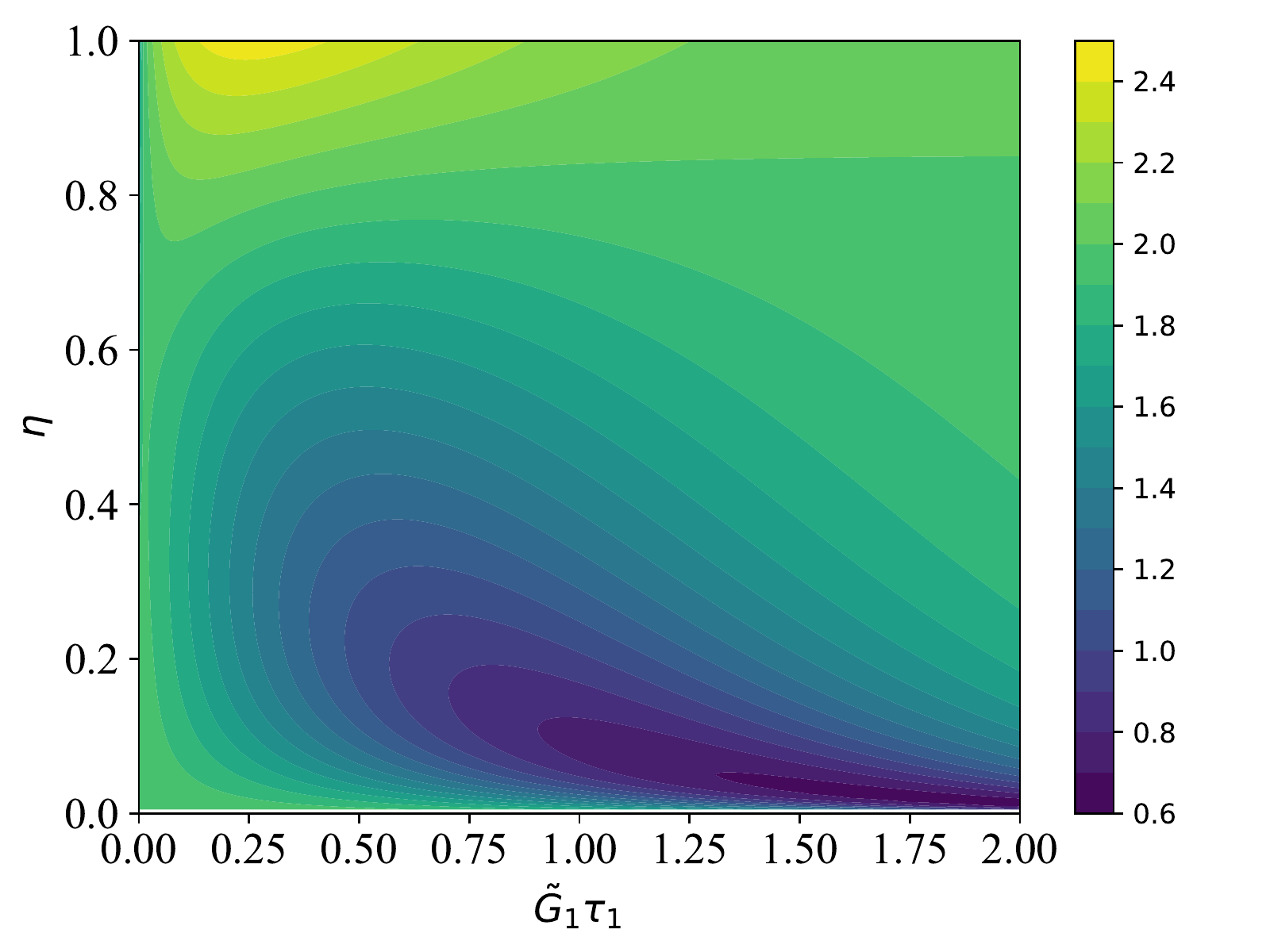}
\caption{(Color online) Contour plot of $S$ versus $\tilde{G}_1\tau_1$ and $\eta$ for the optimal values of $\alpha_{1,2}$ and $\beta_{1,2}$.}
\end{figure}
When the detector efficiency $\eta_d$ and the transmissivity $\lambda_t$ of the beam splitter are considered, we follow Ref. \cite{lee2009testing} and define the overall detection efficiency $\eta=\eta_d\lambda_t$. Thus the CHSH inequality written as a function of $\eta$ becomes \cite{lee2009testing}
\begin{eqnarray}\label{CHSHQ2}
S=&&| \frac{4\pi^2}{\eta^2}[Q_{\eta}({{\alpha_1},{\beta_1}})+Q_{\eta}({{\alpha_1},{\beta_2}})+Q_{\eta}({{\alpha_2},{\beta_1}})\nonumber\\&&-Q_{\eta}({{\alpha_2},{\beta_2}})] \! 
-  \! \frac{4\pi}{\eta} \! [Q_{\eta}(\alpha_1) \! + \! Q_{\eta}(\beta_1)] \!  +  \! 2|\leq 2,
\end{eqnarray}
where the two-mode Q function of the state described in Eq.~(\ref{rhoA1A2}) is given by
\begin{equation} 
Q_{\eta}({{\alpha},{\beta}})=\frac{4}{\pi^2R(\eta)}e^{-2\frac{S(\eta)}{R(\eta)}(|\alpha|^2+|\beta|^2)}e^{\frac{-4\sqrt{p}}{R(\eta)(1-p)}(\alpha^*\beta^*+\alpha\beta)}
\end{equation}
and the single-mode Q function is
\begin{equation}
Q_{\eta}({{\alpha}})=\frac{2}{\pi S(\eta)}e^{-\frac{2}{S(\eta)}|\alpha|^2},
\end{equation}
with $R(\eta)=(1-2/\eta)^2-2(1-2/\eta)(1+p)/(1-p)+1$ and $S(\eta)=(1+p)/(1-p)+2/\eta-1$. Here we have assumed the conversion efficiency $T=1$ for simplicity. 

Figure 4 shows $S$ as a function of $\tilde{G}_1\tau_1$ and $\eta$ at the optimal values of $\alpha_{1,2}$ and $\beta_{1,2}$. Clearly, the violation of Bell inequality requires the detector efficiency $\eta$
larger than $0.8$. As expected, the maximal violation can be obtained at $\eta=1$ and $\tilde{G}_1\tau_1\approx 0.25$. A high overall detection efficiencies can be achieved by the beam splitter with high transmissivity and the photon detector with small dark count probability.

\section{Discussions and Conclusions}
We have assumed the weak coupling condition $G_{1,2} \ll \kappa$ and have neglected the magnon mode decay $\gamma$ in our model. In current YIG-based cavity optomagnonical system, the intrinsic magnon-photon coupling strength has been demonstrated as $g=10.4$ \text{Hz},  the effective coupling strength is obtained as $G=g\alpha=73$ \text{kHz} with $30$ $\mu\text{W}$ optical power, and it can be further enhanced to $G=10$ \text{MHz} \cite{zhang2016optomagnonic}. The weak coupling condition $G_{1,2} \ll \kappa$ is in corresponding with the experimental parameters. However, in order to neglect reasonably the decay of the magnon mode,  the pulse duration should be shorter than the magnon decoherence time, which in turn requires more stronger coupling strength to obtain the optimal value of $\tilde{G}_1\tau_1$ and $\tilde{G}_2\tau_2$. Although the required coupling strength is is not available in optomagnonic system, it is to be expect that the coupling strength will be improved in the future experiments. In addition, we have assumed for simplicity that the magnon mode is initially prepared in its ground state. Indeed, a rigorous proposal should include the case that the magnon mode is initially in a thermal state.  For a YIG sphere with magnon frequency $\omega_m =7.95$ \text{GHz}, the average thermal magnon number in a dilution refrigerator at temperature $10$ \text{mK} is $n_0=0.026$ \cite{lachance2017resolving}. For such a small average thermal magnon number, it may not seriously affect the violation of Bell inequality \cite{vivoli2016proposal,hofer2016proposal}.

In summary, we have proposed a scheme to implement the violation of Bell inequality in cavity optomagnonics, where a magnon mode couples with two nondegenerate cavity modes. Our model is in corresponding with recent YIG-based experiments, and takes into account the selection rules of angular momentum and triple-resonance condition. The Langevin equations of the optomagnonical system are solved and the experimental implementation of the Bell test is analyzed in detail. The results show that a significant violation of Bell inequality can be achieved by a proper choice of $\tilde{G}_1\tau_1$, high magnon-photon conversion efficiency $T$, and high overall detection efficiency $\eta$.

\section*{Acknowledgments}
 We acknowledge supports from the National Natural Science Foundation of China (Grants No.~12174054, No.~12074067, No.~11847056, No.~12004336 and No.~11674059), the Natural Science Foundation of Fujian Province of China (Grants No.~2021J011228, No.~2020J01191 and No.~2019J01431).
\bibliography{opto}

\begin{thebibliography}{67}%
\makeatletter
\providecommand \@ifxundefined [1]{%
 \@ifx{#1\undefined}
}%
\providecommand \@ifnum [1]{%
 \ifnum #1\expandafter \@firstoftwo
 \else \expandafter \@secondoftwo
 \fi
}%
\providecommand \@ifx [1]{%
 \ifx #1\expandafter \@firstoftwo
 \else \expandafter \@secondoftwo
 \fi
}%
\providecommand \natexlab [1]{#1}%
\providecommand \enquote  [1]{``#1''}%
\providecommand \bibnamefont  [1]{#1}%
\providecommand \bibfnamefont [1]{#1}%
\providecommand \citenamefont [1]{#1}%
\providecommand \href@noop [0]{\@secondoftwo}%
\providecommand \href [0]{\begingroup \@sanitize@url \@href}%
\providecommand \@href[1]{\@@startlink{#1}\@@href}%
\providecommand \@@href[1]{\endgroup#1\@@endlink}%
\providecommand \@sanitize@url [0]{\catcode `\\12\catcode `\$12\catcode
  `\&12\catcode `\#12\catcode `\^12\catcode `\_12\catcode `\%12\relax}%
\providecommand \@@startlink[1]{}%
\providecommand \@@endlink[0]{}%
\providecommand \url  [0]{\begingroup\@sanitize@url \@url }%
\providecommand \@url [1]{\endgroup\@href {#1}{\urlprefix }}%
\providecommand \urlprefix  [0]{URL }%
\providecommand \Eprint [0]{\href }%
\providecommand \doibase [0]{https://doi.org/}%
\providecommand \selectlanguage [0]{\@gobble}%
\providecommand \bibinfo  [0]{\@secondoftwo}%
\providecommand \bibfield  [0]{\@secondoftwo}%
\providecommand \translation [1]{[#1]}%
\providecommand \BibitemOpen [0]{}%
\providecommand \bibitemStop [0]{}%
\providecommand \bibitemNoStop [0]{.\EOS\space}%
\providecommand \EOS [0]{\spacefactor3000\relax}%
\providecommand \BibitemShut  [1]{\csname bibitem#1\endcsname}%
\let\auto@bib@innerbib\@empty
\bibitem [{\citenamefont {Duan}\ \emph {et~al.}(2001)\citenamefont {Duan},
  \citenamefont {Lukin}, \citenamefont {Cirac},\ and\ \citenamefont
  {Zoller}}]{duan2001long}%
  \BibitemOpen
  \bibfield  {author} {\bibinfo {author} {\bibfnamefont {L.-M.}\ \bibnamefont
  {Duan}}, \bibinfo {author} {\bibfnamefont {M.~D.}\ \bibnamefont {Lukin}},
  \bibinfo {author} {\bibfnamefont {J.~I.}\ \bibnamefont {Cirac}},\ and\
  \bibinfo {author} {\bibfnamefont {P.}~\bibnamefont {Zoller}},\ }\bibfield
  {title} {\bibinfo {title} {Long-distance quantum communication with atomic
  ensembles and linear optics},\ }\href@noop {} {\bibfield  {journal} {\bibinfo
   {journal} {Nature}\ }\textbf {\bibinfo {volume} {414}},\ \bibinfo {pages}
  {413} (\bibinfo {year} {2001})}\BibitemShut {NoStop}%
\bibitem [{\citenamefont {Kimble}(2008)}]{kimble2008quantum}%
  \BibitemOpen
  \bibfield  {author} {\bibinfo {author} {\bibfnamefont {H.~J.}\ \bibnamefont
  {Kimble}},\ }\bibfield  {title} {\bibinfo {title} {The quantum internet},\
  }\href@noop {} {\bibfield  {journal} {\bibinfo  {journal} {Nature}\ }\textbf
  {\bibinfo {volume} {453}},\ \bibinfo {pages} {1023} (\bibinfo {year}
  {2008})}\BibitemShut {NoStop}%
\bibitem [{\citenamefont {Wehner}\ \emph {et~al.}(2018)\citenamefont {Wehner},
  \citenamefont {Elkouss},\ and\ \citenamefont {Hanson}}]{wehner2018quantum}%
  \BibitemOpen
  \bibfield  {author} {\bibinfo {author} {\bibfnamefont {S.}~\bibnamefont
  {Wehner}}, \bibinfo {author} {\bibfnamefont {D.}~\bibnamefont {Elkouss}},\
  and\ \bibinfo {author} {\bibfnamefont {R.}~\bibnamefont {Hanson}},\
  }\bibfield  {title} {\bibinfo {title} {Quantum internet: A vision for the
  road ahead},\ }\href@noop {} {\bibfield  {journal} {\bibinfo  {journal}
  {Science}\ }\textbf {\bibinfo {volume} {362}} (\bibinfo {year}
  {2018})}\BibitemShut {NoStop}%
\bibitem [{\citenamefont {Degen}\ \emph {et~al.}(2017)\citenamefont {Degen},
  \citenamefont {Reinhard},\ and\ \citenamefont
  {Cappellaro}}]{RevModPhys.89.035002}%
  \BibitemOpen
  \bibfield  {author} {\bibinfo {author} {\bibfnamefont {C.~L.}\ \bibnamefont
  {Degen}}, \bibinfo {author} {\bibfnamefont {F.}~\bibnamefont {Reinhard}},\
  and\ \bibinfo {author} {\bibfnamefont {P.}~\bibnamefont {Cappellaro}},\
  }\bibfield  {title} {\bibinfo {title} {Quantum sensing},\ }\href
  {https://doi.org/10.1103/RevModPhys.89.035002} {\bibfield  {journal}
  {\bibinfo  {journal} {Rev. Mod. Phys.}\ }\textbf {\bibinfo {volume} {89}},\
  \bibinfo {pages} {035002} (\bibinfo {year} {2017})}\BibitemShut {NoStop}%
\bibitem [{\citenamefont {Lachance-Quirion}\ \emph {et~al.}(2019)\citenamefont
  {Lachance-Quirion}, \citenamefont {Tabuchi}, \citenamefont {Gloppe},
  \citenamefont {Usami},\ and\ \citenamefont {Nakamura}}]{lachance2019hybrid}%
  \BibitemOpen
  \bibfield  {author} {\bibinfo {author} {\bibfnamefont {D.}~\bibnamefont
  {Lachance-Quirion}}, \bibinfo {author} {\bibfnamefont {Y.}~\bibnamefont
  {Tabuchi}}, \bibinfo {author} {\bibfnamefont {A.}~\bibnamefont {Gloppe}},
  \bibinfo {author} {\bibfnamefont {K.}~\bibnamefont {Usami}},\ and\ \bibinfo
  {author} {\bibfnamefont {Y.}~\bibnamefont {Nakamura}},\ }\bibfield  {title}
  {\bibinfo {title} {Hybrid quantum systems based on magnonics},\ }\href@noop
  {} {\bibfield  {journal} {\bibinfo  {journal} {Appl. Phys. Express}\ }\textbf
  {\bibinfo {volume} {12}},\ \bibinfo {pages} {070101} (\bibinfo {year}
  {2019})}\BibitemShut {NoStop}%
\bibitem [{\citenamefont {Rameshti}\ \emph {et~al.}(2021)\citenamefont
  {Rameshti}, \citenamefont {Kusminskiy}, \citenamefont {Haigh}, \citenamefont
  {Usami}, \citenamefont {Lachance-Quirion}, \citenamefont {Nakamura},
  \citenamefont {Hu}, \citenamefont {Tang}, \citenamefont {Bauer},\ and\
  \citenamefont {Blanter}}]{rameshti2021cavity}%
  \BibitemOpen
  \bibfield  {author} {\bibinfo {author} {\bibfnamefont {B.~Z.}\ \bibnamefont
  {Rameshti}}, \bibinfo {author} {\bibfnamefont {S.~V.}\ \bibnamefont
  {Kusminskiy}}, \bibinfo {author} {\bibfnamefont {J.~A.}\ \bibnamefont
  {Haigh}}, \bibinfo {author} {\bibfnamefont {K.}~\bibnamefont {Usami}},
  \bibinfo {author} {\bibfnamefont {D.}~\bibnamefont {Lachance-Quirion}},
  \bibinfo {author} {\bibfnamefont {Y.}~\bibnamefont {Nakamura}}, \bibinfo
  {author} {\bibfnamefont {C.-M.}\ \bibnamefont {Hu}}, \bibinfo {author}
  {\bibfnamefont {H.~X.}\ \bibnamefont {Tang}}, \bibinfo {author}
  {\bibfnamefont {G.~E.}\ \bibnamefont {Bauer}},\ and\ \bibinfo {author}
  {\bibfnamefont {Y.~M.}\ \bibnamefont {Blanter}},\ }\bibfield  {title}
  {\bibinfo {title} {Cavity magnonics},\ }\href@noop {} {\bibfield  {journal}
  {\bibinfo  {journal} {arXiv preprint arXiv:2106.09312}\ } (\bibinfo {year}
  {2021})}\BibitemShut {NoStop}%
\bibitem [{\citenamefont {Soykal}\ and\ \citenamefont
  {Flatt\'e}(2010)}]{PhysRevLett.104.077202}%
  \BibitemOpen
  \bibfield  {author} {\bibinfo {author} {\bibfnamefont {O.~O.}\ \bibnamefont
  {Soykal}}\ and\ \bibinfo {author} {\bibfnamefont {M.~E.}\ \bibnamefont
  {Flatt\'e}},\ }\bibfield  {title} {\bibinfo {title} {Strong field
  interactions between a nanomagnet and a photonic cavity},\ }\href
  {https://doi.org/10.1103/PhysRevLett.104.077202} {\bibfield  {journal}
  {\bibinfo  {journal} {Phys. Rev. Lett.}\ }\textbf {\bibinfo {volume} {104}},\
  \bibinfo {pages} {077202} (\bibinfo {year} {2010})}\BibitemShut {NoStop}%
\bibitem [{\citenamefont {Huebl}\ \emph {et~al.}(2013)\citenamefont {Huebl},
  \citenamefont {Zollitsch}, \citenamefont {Lotze}, \citenamefont {Hocke},
  \citenamefont {Greifenstein}, \citenamefont {Marx}, \citenamefont {Gross},\
  and\ \citenamefont {Goennenwein}}]{PhysRevLett.111.127003}%
  \BibitemOpen
  \bibfield  {author} {\bibinfo {author} {\bibfnamefont {H.}~\bibnamefont
  {Huebl}}, \bibinfo {author} {\bibfnamefont {C.~W.}\ \bibnamefont
  {Zollitsch}}, \bibinfo {author} {\bibfnamefont {J.}~\bibnamefont {Lotze}},
  \bibinfo {author} {\bibfnamefont {F.}~\bibnamefont {Hocke}}, \bibinfo
  {author} {\bibfnamefont {M.}~\bibnamefont {Greifenstein}}, \bibinfo {author}
  {\bibfnamefont {A.}~\bibnamefont {Marx}}, \bibinfo {author} {\bibfnamefont
  {R.}~\bibnamefont {Gross}},\ and\ \bibinfo {author} {\bibfnamefont
  {S.~T.~B.}\ \bibnamefont {Goennenwein}},\ }\bibfield  {title} {\bibinfo
  {title} {High cooperativity in coupled microwave resonator ferrimagnetic
  insulator hybrids},\ }\href {https://doi.org/10.1103/PhysRevLett.111.127003}
  {\bibfield  {journal} {\bibinfo  {journal} {Phys. Rev. Lett.}\ }\textbf
  {\bibinfo {volume} {111}},\ \bibinfo {pages} {127003} (\bibinfo {year}
  {2013})}\BibitemShut {NoStop}%
\bibitem [{\citenamefont {Tabuchi}\ \emph {et~al.}(2014)\citenamefont
  {Tabuchi}, \citenamefont {Ishino}, \citenamefont {Ishikawa}, \citenamefont
  {Yamazaki}, \citenamefont {Usami},\ and\ \citenamefont
  {Nakamura}}]{tabuchi2014hybridizing}%
  \BibitemOpen
  \bibfield  {author} {\bibinfo {author} {\bibfnamefont {Y.}~\bibnamefont
  {Tabuchi}}, \bibinfo {author} {\bibfnamefont {S.}~\bibnamefont {Ishino}},
  \bibinfo {author} {\bibfnamefont {T.}~\bibnamefont {Ishikawa}}, \bibinfo
  {author} {\bibfnamefont {R.}~\bibnamefont {Yamazaki}}, \bibinfo {author}
  {\bibfnamefont {K.}~\bibnamefont {Usami}},\ and\ \bibinfo {author}
  {\bibfnamefont {Y.}~\bibnamefont {Nakamura}},\ }\bibfield  {title} {\bibinfo
  {title} {Hybridizing ferromagnetic magnons and microwave photons in the
  quantum limit},\ }\href@noop {} {\bibfield  {journal} {\bibinfo  {journal}
  {Phys. Rev. Lett.}\ }\textbf {\bibinfo {volume} {113}},\ \bibinfo {pages}
  {083603} (\bibinfo {year} {2014})}\BibitemShut {NoStop}%
\bibitem [{\citenamefont {Zhang}\ \emph {et~al.}(2014)\citenamefont {Zhang},
  \citenamefont {Zou}, \citenamefont {Jiang},\ and\ \citenamefont
  {Tang}}]{zhang2014strongly}%
  \BibitemOpen
  \bibfield  {author} {\bibinfo {author} {\bibfnamefont {X.}~\bibnamefont
  {Zhang}}, \bibinfo {author} {\bibfnamefont {C.-L.}\ \bibnamefont {Zou}},
  \bibinfo {author} {\bibfnamefont {L.}~\bibnamefont {Jiang}},\ and\ \bibinfo
  {author} {\bibfnamefont {H.~X.}\ \bibnamefont {Tang}},\ }\bibfield  {title}
  {\bibinfo {title} {Strongly coupled magnons and cavity microwave photons},\
  }\href@noop {} {\bibfield  {journal} {\bibinfo  {journal} {Phys. Rev. Lett.}\
  }\textbf {\bibinfo {volume} {113}},\ \bibinfo {pages} {156401} (\bibinfo
  {year} {2014})}\BibitemShut {NoStop}%
\bibitem [{\citenamefont {Goryachev}\ \emph {et~al.}(2014)\citenamefont
  {Goryachev}, \citenamefont {Farr}, \citenamefont {Creedon}, \citenamefont
  {Fan}, \citenamefont {Kostylev},\ and\ \citenamefont
  {Tobar}}]{PhysRevApplied.2.054002}%
  \BibitemOpen
  \bibfield  {author} {\bibinfo {author} {\bibfnamefont {M.}~\bibnamefont
  {Goryachev}}, \bibinfo {author} {\bibfnamefont {W.~G.}\ \bibnamefont {Farr}},
  \bibinfo {author} {\bibfnamefont {D.~L.}\ \bibnamefont {Creedon}}, \bibinfo
  {author} {\bibfnamefont {Y.}~\bibnamefont {Fan}}, \bibinfo {author}
  {\bibfnamefont {M.}~\bibnamefont {Kostylev}},\ and\ \bibinfo {author}
  {\bibfnamefont {M.~E.}\ \bibnamefont {Tobar}},\ }\bibfield  {title} {\bibinfo
  {title} {High-cooperativity cavity qed with magnons at microwave
  frequencies},\ }\href {https://doi.org/10.1103/PhysRevApplied.2.054002}
  {\bibfield  {journal} {\bibinfo  {journal} {Phys. Rev. Appl.}\ }\textbf
  {\bibinfo {volume} {2}},\ \bibinfo {pages} {054002} (\bibinfo {year}
  {2014})}\BibitemShut {NoStop}%
\bibitem [{\citenamefont {Wang}\ \emph {et~al.}(2018)\citenamefont {Wang},
  \citenamefont {Zhang}, \citenamefont {Zhang}, \citenamefont {Li},
  \citenamefont {Hu},\ and\ \citenamefont {You}}]{PhysRevLett.120.057202}%
  \BibitemOpen
  \bibfield  {author} {\bibinfo {author} {\bibfnamefont {Y.-P.}\ \bibnamefont
  {Wang}}, \bibinfo {author} {\bibfnamefont {G.-Q.}\ \bibnamefont {Zhang}},
  \bibinfo {author} {\bibfnamefont {D.}~\bibnamefont {Zhang}}, \bibinfo
  {author} {\bibfnamefont {T.-F.}\ \bibnamefont {Li}}, \bibinfo {author}
  {\bibfnamefont {C.-M.}\ \bibnamefont {Hu}},\ and\ \bibinfo {author}
  {\bibfnamefont {J.~Q.}\ \bibnamefont {You}},\ }\bibfield  {title} {\bibinfo
  {title} {Bistability of cavity magnon polaritons},\ }\href
  {https://doi.org/10.1103/PhysRevLett.120.057202} {\bibfield  {journal}
  {\bibinfo  {journal} {Phys. Rev. Lett.}\ }\textbf {\bibinfo {volume} {120}},\
  \bibinfo {pages} {057202} (\bibinfo {year} {2018})}\BibitemShut {NoStop}%
\bibitem [{\citenamefont {Hou}\ and\ \citenamefont
  {Liu}(2019)}]{PhysRevLett.123.107702}%
  \BibitemOpen
  \bibfield  {author} {\bibinfo {author} {\bibfnamefont {J.~T.}\ \bibnamefont
  {Hou}}\ and\ \bibinfo {author} {\bibfnamefont {L.}~\bibnamefont {Liu}},\
  }\bibfield  {title} {\bibinfo {title} {Strong coupling between microwave
  photons and nanomagnet magnons},\ }\href
  {https://doi.org/10.1103/PhysRevLett.123.107702} {\bibfield  {journal}
  {\bibinfo  {journal} {Phys. Rev. Lett.}\ }\textbf {\bibinfo {volume} {123}},\
  \bibinfo {pages} {107702} (\bibinfo {year} {2019})}\BibitemShut {NoStop}%
\bibitem [{\citenamefont {Li}\ \emph {et~al.}(2019)\citenamefont {Li},
  \citenamefont {Polakovic}, \citenamefont {Wang}, \citenamefont {Xu},
  \citenamefont {Lendinez}, \citenamefont {Zhang}, \citenamefont {Ding},
  \citenamefont {Khaire}, \citenamefont {Saglam}, \citenamefont {Divan},
  \citenamefont {Pearson}, \citenamefont {Kwok}, \citenamefont {Xiao},
  \citenamefont {Novosad}, \citenamefont {Hoffmann},\ and\ \citenamefont
  {Zhang}}]{PhysRevLett.123.107701}%
  \BibitemOpen
  \bibfield  {author} {\bibinfo {author} {\bibfnamefont {Y.}~\bibnamefont
  {Li}}, \bibinfo {author} {\bibfnamefont {T.}~\bibnamefont {Polakovic}},
  \bibinfo {author} {\bibfnamefont {Y.-L.}\ \bibnamefont {Wang}}, \bibinfo
  {author} {\bibfnamefont {J.}~\bibnamefont {Xu}}, \bibinfo {author}
  {\bibfnamefont {S.}~\bibnamefont {Lendinez}}, \bibinfo {author}
  {\bibfnamefont {Z.}~\bibnamefont {Zhang}}, \bibinfo {author} {\bibfnamefont
  {J.}~\bibnamefont {Ding}}, \bibinfo {author} {\bibfnamefont {T.}~\bibnamefont
  {Khaire}}, \bibinfo {author} {\bibfnamefont {H.}~\bibnamefont {Saglam}},
  \bibinfo {author} {\bibfnamefont {R.}~\bibnamefont {Divan}}, \bibinfo
  {author} {\bibfnamefont {J.}~\bibnamefont {Pearson}}, \bibinfo {author}
  {\bibfnamefont {W.-K.}\ \bibnamefont {Kwok}}, \bibinfo {author}
  {\bibfnamefont {Z.}~\bibnamefont {Xiao}}, \bibinfo {author} {\bibfnamefont
  {V.}~\bibnamefont {Novosad}}, \bibinfo {author} {\bibfnamefont
  {A.}~\bibnamefont {Hoffmann}},\ and\ \bibinfo {author} {\bibfnamefont
  {W.}~\bibnamefont {Zhang}},\ }\bibfield  {title} {\bibinfo {title} {Strong
  coupling between magnons and microwave photons in on-chip
  ferromagnet-superconductor thin-film devices},\ }\href
  {https://doi.org/10.1103/PhysRevLett.123.107701} {\bibfield  {journal}
  {\bibinfo  {journal} {Phys. Rev. Lett.}\ }\textbf {\bibinfo {volume} {123}},\
  \bibinfo {pages} {107701} (\bibinfo {year} {2019})}\BibitemShut {NoStop}%
\bibitem [{\citenamefont {Yuan}\ \emph
  {et~al.}(2020{\natexlab{a}})\citenamefont {Yuan}, \citenamefont {Yan},
  \citenamefont {Zheng}, \citenamefont {He}, \citenamefont {Xia},\ and\
  \citenamefont {Yung}}]{PhysRevLett.124.053602}%
  \BibitemOpen
  \bibfield  {author} {\bibinfo {author} {\bibfnamefont {H.~Y.}\ \bibnamefont
  {Yuan}}, \bibinfo {author} {\bibfnamefont {P.}~\bibnamefont {Yan}}, \bibinfo
  {author} {\bibfnamefont {S.}~\bibnamefont {Zheng}}, \bibinfo {author}
  {\bibfnamefont {Q.~Y.}\ \bibnamefont {He}}, \bibinfo {author} {\bibfnamefont
  {K.}~\bibnamefont {Xia}},\ and\ \bibinfo {author} {\bibfnamefont {M.-H.}\
  \bibnamefont {Yung}},\ }\bibfield  {title} {\bibinfo {title} {Steady bell
  state generation via magnon-photon coupling},\ }\href
  {https://doi.org/10.1103/PhysRevLett.124.053602} {\bibfield  {journal}
  {\bibinfo  {journal} {Phys. Rev. Lett.}\ }\textbf {\bibinfo {volume} {124}},\
  \bibinfo {pages} {053602} (\bibinfo {year} {2020}{\natexlab{a}})}\BibitemShut
  {NoStop}%
\bibitem [{\citenamefont {Yuan}\ \emph
  {et~al.}(2020{\natexlab{b}})\citenamefont {Yuan}, \citenamefont {Zheng},
  \citenamefont {Ficek}, \citenamefont {He},\ and\ \citenamefont
  {Yung}}]{PhysRevB.101.014419}%
  \BibitemOpen
  \bibfield  {author} {\bibinfo {author} {\bibfnamefont {H.~Y.}\ \bibnamefont
  {Yuan}}, \bibinfo {author} {\bibfnamefont {S.}~\bibnamefont {Zheng}},
  \bibinfo {author} {\bibfnamefont {Z.}~\bibnamefont {Ficek}}, \bibinfo
  {author} {\bibfnamefont {Q.~Y.}\ \bibnamefont {He}},\ and\ \bibinfo {author}
  {\bibfnamefont {M.-H.}\ \bibnamefont {Yung}},\ }\bibfield  {title} {\bibinfo
  {title} {Enhancement of magnon-magnon entanglement inside a cavity},\ }\href
  {https://doi.org/10.1103/PhysRevB.101.014419} {\bibfield  {journal} {\bibinfo
   {journal} {Phys. Rev. B}\ }\textbf {\bibinfo {volume} {101}},\ \bibinfo
  {pages} {014419} (\bibinfo {year} {2020}{\natexlab{b}})}\BibitemShut
  {NoStop}%
\bibitem [{\citenamefont {Tabuchi}\ \emph {et~al.}(2015)\citenamefont
  {Tabuchi}, \citenamefont {Ishino}, \citenamefont {Noguchi}, \citenamefont
  {Ishikawa}, \citenamefont {Yamazaki}, \citenamefont {Usami},\ and\
  \citenamefont {Nakamura}}]{tabuchi2015coherent}%
  \BibitemOpen
  \bibfield  {author} {\bibinfo {author} {\bibfnamefont {Y.}~\bibnamefont
  {Tabuchi}}, \bibinfo {author} {\bibfnamefont {S.}~\bibnamefont {Ishino}},
  \bibinfo {author} {\bibfnamefont {A.}~\bibnamefont {Noguchi}}, \bibinfo
  {author} {\bibfnamefont {T.}~\bibnamefont {Ishikawa}}, \bibinfo {author}
  {\bibfnamefont {R.}~\bibnamefont {Yamazaki}}, \bibinfo {author}
  {\bibfnamefont {K.}~\bibnamefont {Usami}},\ and\ \bibinfo {author}
  {\bibfnamefont {Y.}~\bibnamefont {Nakamura}},\ }\bibfield  {title} {\bibinfo
  {title} {Coherent coupling between a ferromagnetic magnon and a
  superconducting qubit},\ }\href@noop {} {\bibfield  {journal} {\bibinfo
  {journal} {Science}\ }\textbf {\bibinfo {volume} {349}},\ \bibinfo {pages}
  {405} (\bibinfo {year} {2015})}\BibitemShut {NoStop}%
\bibitem [{\citenamefont {Lachance-Quirion}\ \emph {et~al.}(2017)\citenamefont
  {Lachance-Quirion}, \citenamefont {Tabuchi}, \citenamefont {Ishino},
  \citenamefont {Noguchi}, \citenamefont {Ishikawa}, \citenamefont {Yamazaki},\
  and\ \citenamefont {Nakamura}}]{lachance2017resolving}%
  \BibitemOpen
  \bibfield  {author} {\bibinfo {author} {\bibfnamefont {D.}~\bibnamefont
  {Lachance-Quirion}}, \bibinfo {author} {\bibfnamefont {Y.}~\bibnamefont
  {Tabuchi}}, \bibinfo {author} {\bibfnamefont {S.}~\bibnamefont {Ishino}},
  \bibinfo {author} {\bibfnamefont {A.}~\bibnamefont {Noguchi}}, \bibinfo
  {author} {\bibfnamefont {T.}~\bibnamefont {Ishikawa}}, \bibinfo {author}
  {\bibfnamefont {R.}~\bibnamefont {Yamazaki}},\ and\ \bibinfo {author}
  {\bibfnamefont {Y.}~\bibnamefont {Nakamura}},\ }\bibfield  {title} {\bibinfo
  {title} {Resolving quanta of collective spin excitations in a
  millimeter-sized ferromagnet},\ }\href@noop {} {\bibfield  {journal}
  {\bibinfo  {journal} {Sci. Adv.}\ }\textbf {\bibinfo {volume} {3}},\ \bibinfo
  {pages} {e1603150} (\bibinfo {year} {2017})}\BibitemShut {NoStop}%
\bibitem [{\citenamefont {Zhang}\ \emph
  {et~al.}(2016{\natexlab{a}})\citenamefont {Zhang}, \citenamefont {Zou},
  \citenamefont {Jiang},\ and\ \citenamefont {Tang}}]{zhang2016cavity}%
  \BibitemOpen
  \bibfield  {author} {\bibinfo {author} {\bibfnamefont {X.}~\bibnamefont
  {Zhang}}, \bibinfo {author} {\bibfnamefont {C.-L.}\ \bibnamefont {Zou}},
  \bibinfo {author} {\bibfnamefont {L.}~\bibnamefont {Jiang}},\ and\ \bibinfo
  {author} {\bibfnamefont {H.~X.}\ \bibnamefont {Tang}},\ }\bibfield  {title}
  {\bibinfo {title} {Cavity magnomechanics},\ }\href@noop {} {\bibfield
  {journal} {\bibinfo  {journal} {Sci. Adv.}\ }\textbf {\bibinfo {volume}
  {2}},\ \bibinfo {pages} {e1501286} (\bibinfo {year}
  {2016}{\natexlab{a}})}\BibitemShut {NoStop}%
\bibitem [{\citenamefont {Li}\ \emph {et~al.}(2018)\citenamefont {Li},
  \citenamefont {Zhu},\ and\ \citenamefont {Agarwal}}]{PhysRevLett.121.203601}%
  \BibitemOpen
  \bibfield  {author} {\bibinfo {author} {\bibfnamefont {J.}~\bibnamefont
  {Li}}, \bibinfo {author} {\bibfnamefont {S.-Y.}\ \bibnamefont {Zhu}},\ and\
  \bibinfo {author} {\bibfnamefont {G.~S.}\ \bibnamefont {Agarwal}},\
  }\bibfield  {title} {\bibinfo {title} {Magnon-photon-phonon entanglement in
  cavity magnomechanics},\ }\href
  {https://doi.org/10.1103/PhysRevLett.121.203601} {\bibfield  {journal}
  {\bibinfo  {journal} {Phys. Rev. Lett.}\ }\textbf {\bibinfo {volume} {121}},\
  \bibinfo {pages} {203601} (\bibinfo {year} {2018})}\BibitemShut {NoStop}%
\bibitem [{\citenamefont {Viola~Kusminskiy}(2019)}]{kusminskiy2019cavity}%
  \BibitemOpen
  \bibfield  {author} {\bibinfo {author} {\bibfnamefont {S.}~\bibnamefont
  {Viola~Kusminskiy}},\ }\bibfield  {title} {\bibinfo {title} {Cavity
  optomagnonics},\ }\href@noop {} {\bibfield  {journal} {\bibinfo  {journal}
  {arXiv preprint arXiv:1911.11104}\ } (\bibinfo {year} {2019})}\BibitemShut
  {NoStop}%
\bibitem [{\citenamefont {Hisatomi}\ \emph {et~al.}(2016)\citenamefont
  {Hisatomi}, \citenamefont {Osada}, \citenamefont {Tabuchi}, \citenamefont
  {Ishikawa}, \citenamefont {Noguchi}, \citenamefont {Yamazaki}, \citenamefont
  {Usami},\ and\ \citenamefont {Nakamura}}]{hisatomi2016bidirectional}%
  \BibitemOpen
  \bibfield  {author} {\bibinfo {author} {\bibfnamefont {R.}~\bibnamefont
  {Hisatomi}}, \bibinfo {author} {\bibfnamefont {A.}~\bibnamefont {Osada}},
  \bibinfo {author} {\bibfnamefont {Y.}~\bibnamefont {Tabuchi}}, \bibinfo
  {author} {\bibfnamefont {T.}~\bibnamefont {Ishikawa}}, \bibinfo {author}
  {\bibfnamefont {A.}~\bibnamefont {Noguchi}}, \bibinfo {author} {\bibfnamefont
  {R.}~\bibnamefont {Yamazaki}}, \bibinfo {author} {\bibfnamefont
  {K.}~\bibnamefont {Usami}},\ and\ \bibinfo {author} {\bibfnamefont
  {Y.}~\bibnamefont {Nakamura}},\ }\bibfield  {title} {\bibinfo {title}
  {Bidirectional conversion between microwave and light via ferromagnetic
  magnons},\ }\href@noop {} {\bibfield  {journal} {\bibinfo  {journal} {Phys.
  Rev. B}\ }\textbf {\bibinfo {volume} {93}},\ \bibinfo {pages} {174427}
  (\bibinfo {year} {2016})}\BibitemShut {NoStop}%
\bibitem [{\citenamefont {Osada}\ \emph {et~al.}(2016)\citenamefont {Osada},
  \citenamefont {Hisatomi}, \citenamefont {Noguchi}, \citenamefont {Tabuchi},
  \citenamefont {Yamazaki}, \citenamefont {Usami}, \citenamefont {Sadgrove},
  \citenamefont {Yalla}, \citenamefont {Nomura},\ and\ \citenamefont
  {Nakamura}}]{osada2016cavity}%
  \BibitemOpen
  \bibfield  {author} {\bibinfo {author} {\bibfnamefont {A.}~\bibnamefont
  {Osada}}, \bibinfo {author} {\bibfnamefont {R.}~\bibnamefont {Hisatomi}},
  \bibinfo {author} {\bibfnamefont {A.}~\bibnamefont {Noguchi}}, \bibinfo
  {author} {\bibfnamefont {Y.}~\bibnamefont {Tabuchi}}, \bibinfo {author}
  {\bibfnamefont {R.}~\bibnamefont {Yamazaki}}, \bibinfo {author}
  {\bibfnamefont {K.}~\bibnamefont {Usami}}, \bibinfo {author} {\bibfnamefont
  {M.}~\bibnamefont {Sadgrove}}, \bibinfo {author} {\bibfnamefont
  {R.}~\bibnamefont {Yalla}}, \bibinfo {author} {\bibfnamefont
  {M.}~\bibnamefont {Nomura}},\ and\ \bibinfo {author} {\bibfnamefont
  {Y.}~\bibnamefont {Nakamura}},\ }\bibfield  {title} {\bibinfo {title} {Cavity
  optomagnonics with spin-orbit coupled photons},\ }\href@noop {} {\bibfield
  {journal} {\bibinfo  {journal} {Phys. Rev. Lett.}\ }\textbf {\bibinfo
  {volume} {116}},\ \bibinfo {pages} {223601} (\bibinfo {year}
  {2016})}\BibitemShut {NoStop}%
\bibitem [{\citenamefont {Zhang}\ \emph
  {et~al.}(2016{\natexlab{b}})\citenamefont {Zhang}, \citenamefont {Zhu},
  \citenamefont {Zou},\ and\ \citenamefont {Tang}}]{zhang2016optomagnonic}%
  \BibitemOpen
  \bibfield  {author} {\bibinfo {author} {\bibfnamefont {X.}~\bibnamefont
  {Zhang}}, \bibinfo {author} {\bibfnamefont {N.}~\bibnamefont {Zhu}}, \bibinfo
  {author} {\bibfnamefont {C.-L.}\ \bibnamefont {Zou}},\ and\ \bibinfo {author}
  {\bibfnamefont {H.~X.}\ \bibnamefont {Tang}},\ }\bibfield  {title} {\bibinfo
  {title} {Optomagnonic whispering gallery microresonators},\ }\href@noop {}
  {\bibfield  {journal} {\bibinfo  {journal} {Phys. Rev. Lett.}\ }\textbf
  {\bibinfo {volume} {117}},\ \bibinfo {pages} {123605} (\bibinfo {year}
  {2016}{\natexlab{b}})}\BibitemShut {NoStop}%
\bibitem [{\citenamefont {Haigh}\ \emph {et~al.}(2016)\citenamefont {Haigh},
  \citenamefont {Nunnenkamp}, \citenamefont {Ramsay},\ and\ \citenamefont
  {Ferguson}}]{haigh2016triple}%
  \BibitemOpen
  \bibfield  {author} {\bibinfo {author} {\bibfnamefont {J.~A.}\ \bibnamefont
  {Haigh}}, \bibinfo {author} {\bibfnamefont {A.}~\bibnamefont {Nunnenkamp}},
  \bibinfo {author} {\bibfnamefont {A.~J.}\ \bibnamefont {Ramsay}},\ and\
  \bibinfo {author} {\bibfnamefont {A.~J.}\ \bibnamefont {Ferguson}},\
  }\bibfield  {title} {\bibinfo {title} {Triple-resonant brillouin light
  scattering in magneto-optical cavities},\ }\href@noop {} {\bibfield
  {journal} {\bibinfo  {journal} {Phys. Rev. Lett.}\ }\textbf {\bibinfo
  {volume} {117}},\ \bibinfo {pages} {133602} (\bibinfo {year}
  {2016})}\BibitemShut {NoStop}%
\bibitem [{\citenamefont {Osada}\ \emph
  {et~al.}(2018{\natexlab{a}})\citenamefont {Osada}, \citenamefont {Gloppe},
  \citenamefont {Hisatomi}, \citenamefont {Noguchi}, \citenamefont {Yamazaki},
  \citenamefont {Nomura}, \citenamefont {Nakamura},\ and\ \citenamefont
  {Usami}}]{PhysRevLett.120.133602}%
  \BibitemOpen
  \bibfield  {author} {\bibinfo {author} {\bibfnamefont {A.}~\bibnamefont
  {Osada}}, \bibinfo {author} {\bibfnamefont {A.}~\bibnamefont {Gloppe}},
  \bibinfo {author} {\bibfnamefont {R.}~\bibnamefont {Hisatomi}}, \bibinfo
  {author} {\bibfnamefont {A.}~\bibnamefont {Noguchi}}, \bibinfo {author}
  {\bibfnamefont {R.}~\bibnamefont {Yamazaki}}, \bibinfo {author}
  {\bibfnamefont {M.}~\bibnamefont {Nomura}}, \bibinfo {author} {\bibfnamefont
  {Y.}~\bibnamefont {Nakamura}},\ and\ \bibinfo {author} {\bibfnamefont
  {K.}~\bibnamefont {Usami}},\ }\bibfield  {title} {\bibinfo {title} {Brillouin
  light scattering by magnetic quasivortices in cavity optomagnonics},\ }\href
  {https://doi.org/10.1103/PhysRevLett.120.133602} {\bibfield  {journal}
  {\bibinfo  {journal} {Phys. Rev. Lett.}\ }\textbf {\bibinfo {volume} {120}},\
  \bibinfo {pages} {133602} (\bibinfo {year} {2018}{\natexlab{a}})}\BibitemShut
  {NoStop}%
\bibitem [{\citenamefont {Hisatomi}\ \emph {et~al.}(2019)\citenamefont
  {Hisatomi}, \citenamefont {Noguchi}, \citenamefont {Yamazaki}, \citenamefont
  {Nakata}, \citenamefont {Gloppe}, \citenamefont {Nakamura},\ and\
  \citenamefont {Usami}}]{PhysRevLett.123.207401}%
  \BibitemOpen
  \bibfield  {author} {\bibinfo {author} {\bibfnamefont {R.}~\bibnamefont
  {Hisatomi}}, \bibinfo {author} {\bibfnamefont {A.}~\bibnamefont {Noguchi}},
  \bibinfo {author} {\bibfnamefont {R.}~\bibnamefont {Yamazaki}}, \bibinfo
  {author} {\bibfnamefont {Y.}~\bibnamefont {Nakata}}, \bibinfo {author}
  {\bibfnamefont {A.}~\bibnamefont {Gloppe}}, \bibinfo {author} {\bibfnamefont
  {Y.}~\bibnamefont {Nakamura}},\ and\ \bibinfo {author} {\bibfnamefont
  {K.}~\bibnamefont {Usami}},\ }\bibfield  {title} {\bibinfo {title}
  {Helicity-changing brillouin light scattering by magnons in a ferromagnetic
  crystal},\ }\href {https://doi.org/10.1103/PhysRevLett.123.207401} {\bibfield
   {journal} {\bibinfo  {journal} {Phys. Rev. Lett.}\ }\textbf {\bibinfo
  {volume} {123}},\ \bibinfo {pages} {207401} (\bibinfo {year}
  {2019})}\BibitemShut {NoStop}%
\bibitem [{\citenamefont {Liu}\ \emph {et~al.}(2016)\citenamefont {Liu},
  \citenamefont {Zhang}, \citenamefont {Tang},\ and\ \citenamefont
  {Flatt{\'e}}}]{liu2016optomagnonics}%
  \BibitemOpen
  \bibfield  {author} {\bibinfo {author} {\bibfnamefont {T.}~\bibnamefont
  {Liu}}, \bibinfo {author} {\bibfnamefont {X.}~\bibnamefont {Zhang}}, \bibinfo
  {author} {\bibfnamefont {H.~X.}\ \bibnamefont {Tang}},\ and\ \bibinfo
  {author} {\bibfnamefont {M.~E.}\ \bibnamefont {Flatt{\'e}}},\ }\bibfield
  {title} {\bibinfo {title} {Optomagnonics in magnetic solids},\ }\href@noop {}
  {\bibfield  {journal} {\bibinfo  {journal} {Phys. Rev. B}\ }\textbf {\bibinfo
  {volume} {94}},\ \bibinfo {pages} {060405(R)} (\bibinfo {year}
  {2016})}\BibitemShut {NoStop}%
\bibitem [{\citenamefont {Viola~Kusminskiy}\ \emph {et~al.}(2016)\citenamefont
  {Viola~Kusminskiy}, \citenamefont {Tang},\ and\ \citenamefont
  {Marquardt}}]{kusminskiy2016coupled}%
  \BibitemOpen
  \bibfield  {author} {\bibinfo {author} {\bibfnamefont {S.}~\bibnamefont
  {Viola~Kusminskiy}}, \bibinfo {author} {\bibfnamefont {H.~X.}\ \bibnamefont
  {Tang}},\ and\ \bibinfo {author} {\bibfnamefont {F.}~\bibnamefont
  {Marquardt}},\ }\bibfield  {title} {\bibinfo {title} {Coupled spin-light
  dynamics in cavity optomagnonics},\ }\href@noop {} {\bibfield  {journal}
  {\bibinfo  {journal} {Phys. Rev. A}\ }\textbf {\bibinfo {volume} {94}},\
  \bibinfo {pages} {033821} (\bibinfo {year} {2016})}\BibitemShut {NoStop}%
\bibitem [{\citenamefont {Sharma}\ \emph {et~al.}(2017)\citenamefont {Sharma},
  \citenamefont {Blanter},\ and\ \citenamefont {Bauer}}]{sharma2017light}%
  \BibitemOpen
  \bibfield  {author} {\bibinfo {author} {\bibfnamefont {S.}~\bibnamefont
  {Sharma}}, \bibinfo {author} {\bibfnamefont {Y.~M.}\ \bibnamefont
  {Blanter}},\ and\ \bibinfo {author} {\bibfnamefont {G.~E.~W.}\ \bibnamefont
  {Bauer}},\ }\bibfield  {title} {\bibinfo {title} {Light scattering by magnons
  in whispering gallery mode cavities},\ }\href@noop {} {\bibfield  {journal}
  {\bibinfo  {journal} {Phys. Rev. B}\ }\textbf {\bibinfo {volume} {96}},\
  \bibinfo {pages} {094412} (\bibinfo {year} {2017})}\BibitemShut {NoStop}%
\bibitem [{\citenamefont {Graf}\ \emph {et~al.}(2018)\citenamefont {Graf},
  \citenamefont {Pfeifer}, \citenamefont {Marquardt},\ and\ \citenamefont
  {Viola~Kusminskiy}}]{graf2018cavity}%
  \BibitemOpen
  \bibfield  {author} {\bibinfo {author} {\bibfnamefont {J.}~\bibnamefont
  {Graf}}, \bibinfo {author} {\bibfnamefont {H.}~\bibnamefont {Pfeifer}},
  \bibinfo {author} {\bibfnamefont {F.}~\bibnamefont {Marquardt}},\ and\
  \bibinfo {author} {\bibfnamefont {S.}~\bibnamefont {Viola~Kusminskiy}},\
  }\bibfield  {title} {\bibinfo {title} {Cavity optomagnonics with magnetic
  textures: Coupling a magnetic vortex to light},\ }\href@noop {} {\bibfield
  {journal} {\bibinfo  {journal} {Phys. Rev. B}\ }\textbf {\bibinfo {volume}
  {98}},\ \bibinfo {pages} {241406(R)} (\bibinfo {year} {2018})}\BibitemShut
  {NoStop}%
\bibitem [{\citenamefont {Osada}\ \emph
  {et~al.}(2018{\natexlab{b}})\citenamefont {Osada}, \citenamefont {Gloppe},
  \citenamefont {Nakamura},\ and\ \citenamefont {Usami}}]{osada2018orbital}%
  \BibitemOpen
  \bibfield  {author} {\bibinfo {author} {\bibfnamefont {A.}~\bibnamefont
  {Osada}}, \bibinfo {author} {\bibfnamefont {A.}~\bibnamefont {Gloppe}},
  \bibinfo {author} {\bibfnamefont {Y.}~\bibnamefont {Nakamura}},\ and\
  \bibinfo {author} {\bibfnamefont {K.}~\bibnamefont {Usami}},\ }\bibfield
  {title} {\bibinfo {title} {Orbital angular momentum conservation in brillouin
  light scattering within a ferromagnetic sphere},\ }\href@noop {} {\bibfield
  {journal} {\bibinfo  {journal} {New J. Phys.}\ }\textbf {\bibinfo {volume}
  {20}},\ \bibinfo {pages} {103018} (\bibinfo {year}
  {2018}{\natexlab{b}})}\BibitemShut {NoStop}%
\bibitem [{\citenamefont {Haigh}\ \emph {et~al.}(2018)\citenamefont {Haigh},
  \citenamefont {Lambert}, \citenamefont {Sharma}, \citenamefont {Blanter},
  \citenamefont {Bauer},\ and\ \citenamefont {Ramsay}}]{haigh2018selection}%
  \BibitemOpen
  \bibfield  {author} {\bibinfo {author} {\bibfnamefont {J.~A.}\ \bibnamefont
  {Haigh}}, \bibinfo {author} {\bibfnamefont {N.~J.}\ \bibnamefont {Lambert}},
  \bibinfo {author} {\bibfnamefont {S.}~\bibnamefont {Sharma}}, \bibinfo
  {author} {\bibfnamefont {Y.~M.}\ \bibnamefont {Blanter}}, \bibinfo {author}
  {\bibfnamefont {G.~E.~W.}\ \bibnamefont {Bauer}},\ and\ \bibinfo {author}
  {\bibfnamefont {A.~J.}\ \bibnamefont {Ramsay}},\ }\bibfield  {title}
  {\bibinfo {title} {Selection rules for cavity-enhanced brillouin light
  scattering from magnetostatic modes},\ }\href@noop {} {\bibfield  {journal}
  {\bibinfo  {journal} {Phys. Rev. B}\ }\textbf {\bibinfo {volume} {97}},\
  \bibinfo {pages} {214423} (\bibinfo {year} {2018})}\BibitemShut {NoStop}%
\bibitem [{\citenamefont {Sharma}\ \emph {et~al.}(2019)\citenamefont {Sharma},
  \citenamefont {Rameshti}, \citenamefont {Blanter},\ and\ \citenamefont
  {Bauer}}]{sharma2019optimal}%
  \BibitemOpen
  \bibfield  {author} {\bibinfo {author} {\bibfnamefont {S.}~\bibnamefont
  {Sharma}}, \bibinfo {author} {\bibfnamefont {B.~Z.}\ \bibnamefont
  {Rameshti}}, \bibinfo {author} {\bibfnamefont {Y.~M.}\ \bibnamefont
  {Blanter}},\ and\ \bibinfo {author} {\bibfnamefont {G.~E.~W.}\ \bibnamefont
  {Bauer}},\ }\bibfield  {title} {\bibinfo {title} {Optimal mode matching in
  cavity optomagnonics},\ }\href@noop {} {\bibfield  {journal} {\bibinfo
  {journal} {Phys. Rev. B}\ }\textbf {\bibinfo {volume} {99}},\ \bibinfo
  {pages} {214423} (\bibinfo {year} {2019})}\BibitemShut {NoStop}%
\bibitem [{\citenamefont {Wu}\ \emph {et~al.}(2021)\citenamefont {Wu},
  \citenamefont {Wang}, \citenamefont {Wu}, \citenamefont {Li},\ and\
  \citenamefont {You}}]{PhysRevA.104.023711}%
  \BibitemOpen
  \bibfield  {author} {\bibinfo {author} {\bibfnamefont {W.-J.}\ \bibnamefont
  {Wu}}, \bibinfo {author} {\bibfnamefont {Y.-P.}\ \bibnamefont {Wang}},
  \bibinfo {author} {\bibfnamefont {J.-Z.}\ \bibnamefont {Wu}}, \bibinfo
  {author} {\bibfnamefont {J.}~\bibnamefont {Li}},\ and\ \bibinfo {author}
  {\bibfnamefont {J.~Q.}\ \bibnamefont {You}},\ }\bibfield  {title} {\bibinfo
  {title} {Remote magnon entanglement between two massive ferrimagnetic spheres
  via cavity optomagnonics},\ }\href
  {https://doi.org/10.1103/PhysRevA.104.023711} {\bibfield  {journal} {\bibinfo
   {journal} {Phys. Rev. A}\ }\textbf {\bibinfo {volume} {104}},\ \bibinfo
  {pages} {023711} (\bibinfo {year} {2021})}\BibitemShut {NoStop}%
\bibitem [{\citenamefont {Sharma}\ \emph {et~al.}(2018)\citenamefont {Sharma},
  \citenamefont {Blanter},\ and\ \citenamefont {Bauer}}]{sharma2018optical}%
  \BibitemOpen
  \bibfield  {author} {\bibinfo {author} {\bibfnamefont {S.}~\bibnamefont
  {Sharma}}, \bibinfo {author} {\bibfnamefont {Y.~M.}\ \bibnamefont
  {Blanter}},\ and\ \bibinfo {author} {\bibfnamefont {G.~E.~W.}\ \bibnamefont
  {Bauer}},\ }\bibfield  {title} {\bibinfo {title} {Optical cooling of
  magnons},\ }\href@noop {} {\bibfield  {journal} {\bibinfo  {journal} {Phys.
  Rev. Lett.}\ }\textbf {\bibinfo {volume} {121}},\ \bibinfo {pages} {087205}
  (\bibinfo {year} {2018})}\BibitemShut {NoStop}%
\bibitem [{\citenamefont {Bittencourt}\ \emph {et~al.}(2019)\citenamefont
  {Bittencourt}, \citenamefont {Feulner},\ and\ \citenamefont
  {Kusminskiy}}]{bittencourt2019magnon}%
  \BibitemOpen
  \bibfield  {author} {\bibinfo {author} {\bibfnamefont {V.~A. S.~V.}\
  \bibnamefont {Bittencourt}}, \bibinfo {author} {\bibfnamefont
  {V.}~\bibnamefont {Feulner}},\ and\ \bibinfo {author} {\bibfnamefont {S.~V.}\
  \bibnamefont {Kusminskiy}},\ }\bibfield  {title} {\bibinfo {title} {Magnon
  heralding in cavity optomagnonics},\ }\href@noop {} {\bibfield  {journal}
  {\bibinfo  {journal} {Phys. Rev. A}\ }\textbf {\bibinfo {volume} {100}},\
  \bibinfo {pages} {013810} (\bibinfo {year} {2019})}\BibitemShut {NoStop}%
\bibitem [{\citenamefont {Gao}\ \emph {et~al.}(2019)\citenamefont {Gao},
  \citenamefont {Liu}, \citenamefont {Wang}, \citenamefont {Cao},\ and\
  \citenamefont {Wang}}]{PhysRevA.100.043831}%
  \BibitemOpen
  \bibfield  {author} {\bibinfo {author} {\bibfnamefont {Y.-P.}\ \bibnamefont
  {Gao}}, \bibinfo {author} {\bibfnamefont {X.-F.}\ \bibnamefont {Liu}},
  \bibinfo {author} {\bibfnamefont {T.-J.}\ \bibnamefont {Wang}}, \bibinfo
  {author} {\bibfnamefont {C.}~\bibnamefont {Cao}},\ and\ \bibinfo {author}
  {\bibfnamefont {C.}~\bibnamefont {Wang}},\ }\bibfield  {title} {\bibinfo
  {title} {Photon excitation and photon-blockade effects in optomagnonic
  microcavities},\ }\href {https://doi.org/10.1103/PhysRevA.100.043831}
  {\bibfield  {journal} {\bibinfo  {journal} {Phys. Rev. A}\ }\textbf {\bibinfo
  {volume} {100}},\ \bibinfo {pages} {043831} (\bibinfo {year}
  {2019})}\BibitemShut {NoStop}%
\bibitem [{\citenamefont {Sharma}\ \emph {et~al.}(2021)\citenamefont {Sharma},
  \citenamefont {Bittencourt}, \citenamefont {Karenowska},\ and\ \citenamefont
  {Kusminskiy}}]{PhysRevB.103.L100403}%
  \BibitemOpen
  \bibfield  {author} {\bibinfo {author} {\bibfnamefont {S.}~\bibnamefont
  {Sharma}}, \bibinfo {author} {\bibfnamefont {V.~A. S.~V.}\ \bibnamefont
  {Bittencourt}}, \bibinfo {author} {\bibfnamefont {A.~D.}\ \bibnamefont
  {Karenowska}},\ and\ \bibinfo {author} {\bibfnamefont {S.~V.}\ \bibnamefont
  {Kusminskiy}},\ }\bibfield  {title} {\bibinfo {title} {Spin cat states in
  ferromagnetic insulators},\ }\href
  {https://doi.org/10.1103/PhysRevB.103.L100403} {\bibfield  {journal}
  {\bibinfo  {journal} {Phys. Rev. B}\ }\textbf {\bibinfo {volume} {103}},\
  \bibinfo {pages} {L100403} (\bibinfo {year} {2021})}\BibitemShut {NoStop}%
\bibitem [{\citenamefont {Sun}\ \emph {et~al.}(2021)\citenamefont {Sun},
  \citenamefont {Zheng}, \citenamefont {Xiao}, \citenamefont {Gong},
  \citenamefont {He},\ and\ \citenamefont {Xia}}]{PhysRevLett.127.087203}%
  \BibitemOpen
  \bibfield  {author} {\bibinfo {author} {\bibfnamefont {F.-X.}\ \bibnamefont
  {Sun}}, \bibinfo {author} {\bibfnamefont {S.-S.}\ \bibnamefont {Zheng}},
  \bibinfo {author} {\bibfnamefont {Y.}~\bibnamefont {Xiao}}, \bibinfo {author}
  {\bibfnamefont {Q.}~\bibnamefont {Gong}}, \bibinfo {author} {\bibfnamefont
  {Q.}~\bibnamefont {He}},\ and\ \bibinfo {author} {\bibfnamefont
  {K.}~\bibnamefont {Xia}},\ }\bibfield  {title} {\bibinfo {title} {Remote
  generation of magnon schr\"odinger cat state via magnon-photon
  entanglement},\ }\href {https://doi.org/10.1103/PhysRevLett.127.087203}
  {\bibfield  {journal} {\bibinfo  {journal} {Phys. Rev. Lett.}\ }\textbf
  {\bibinfo {volume} {127}},\ \bibinfo {pages} {087203} (\bibinfo {year}
  {2021})}\BibitemShut {NoStop}%
\bibitem [{\citenamefont {Brunner}\ \emph {et~al.}(2014)\citenamefont
  {Brunner}, \citenamefont {Cavalcanti}, \citenamefont {Pironio}, \citenamefont
  {Scarani},\ and\ \citenamefont {Wehner}}]{RevModPhys.86.419}%
  \BibitemOpen
  \bibfield  {author} {\bibinfo {author} {\bibfnamefont {N.}~\bibnamefont
  {Brunner}}, \bibinfo {author} {\bibfnamefont {D.}~\bibnamefont {Cavalcanti}},
  \bibinfo {author} {\bibfnamefont {S.}~\bibnamefont {Pironio}}, \bibinfo
  {author} {\bibfnamefont {V.}~\bibnamefont {Scarani}},\ and\ \bibinfo {author}
  {\bibfnamefont {S.}~\bibnamefont {Wehner}},\ }\bibfield  {title} {\bibinfo
  {title} {Bell nonlocality},\ }\href
  {https://doi.org/10.1103/RevModPhys.86.419} {\bibfield  {journal} {\bibinfo
  {journal} {Rev. Mod. Phys.}\ }\textbf {\bibinfo {volume} {86}},\ \bibinfo
  {pages} {419} (\bibinfo {year} {2014})}\BibitemShut {NoStop}%
\bibitem [{\citenamefont {Clauser}\ \emph {et~al.}(1969)\citenamefont
  {Clauser}, \citenamefont {Horne}, \citenamefont {Shimony},\ and\
  \citenamefont {Holt}}]{PhysRevLett.23.880}%
  \BibitemOpen
  \bibfield  {author} {\bibinfo {author} {\bibfnamefont {J.~F.}\ \bibnamefont
  {Clauser}}, \bibinfo {author} {\bibfnamefont {M.~A.}\ \bibnamefont {Horne}},
  \bibinfo {author} {\bibfnamefont {A.}~\bibnamefont {Shimony}},\ and\ \bibinfo
  {author} {\bibfnamefont {R.~A.}\ \bibnamefont {Holt}},\ }\bibfield  {title}
  {\bibinfo {title} {Proposed experiment to test local hidden-variable
  theories},\ }\href {https://doi.org/10.1103/PhysRevLett.23.880} {\bibfield
  {journal} {\bibinfo  {journal} {Phys. Rev. Lett.}\ }\textbf {\bibinfo
  {volume} {23}},\ \bibinfo {pages} {880} (\bibinfo {year} {1969})}\BibitemShut
  {NoStop}%
\bibitem [{\citenamefont {Freedman}\ and\ \citenamefont
  {Clauser}(1972)}]{freedman1972experimental}%
  \BibitemOpen
  \bibfield  {author} {\bibinfo {author} {\bibfnamefont {S.~J.}\ \bibnamefont
  {Freedman}}\ and\ \bibinfo {author} {\bibfnamefont {J.~F.}\ \bibnamefont
  {Clauser}},\ }\bibfield  {title} {\bibinfo {title} {Experimental test of
  local hidden-variable theories},\ }\href@noop {} {\bibfield  {journal}
  {\bibinfo  {journal} {Phys. Rev. Lett.}\ }\textbf {\bibinfo {volume} {28}},\
  \bibinfo {pages} {938} (\bibinfo {year} {1972})}\BibitemShut {NoStop}%
\bibitem [{\citenamefont {Aspect}\ \emph {et~al.}(1981)\citenamefont {Aspect},
  \citenamefont {Grangier},\ and\ \citenamefont
  {Roger}}]{aspect1981experimental}%
  \BibitemOpen
  \bibfield  {author} {\bibinfo {author} {\bibfnamefont {A.}~\bibnamefont
  {Aspect}}, \bibinfo {author} {\bibfnamefont {P.}~\bibnamefont {Grangier}},\
  and\ \bibinfo {author} {\bibfnamefont {G.}~\bibnamefont {Roger}},\ }\bibfield
   {title} {\bibinfo {title} {Experimental tests of realistic local theories
  via bell's theorem},\ }\href@noop {} {\bibfield  {journal} {\bibinfo
  {journal} {Phys. Rev. Lett.}\ }\textbf {\bibinfo {volume} {47}},\ \bibinfo
  {pages} {460} (\bibinfo {year} {1981})}\BibitemShut {NoStop}%
\bibitem [{\citenamefont {Shih}\ and\ \citenamefont
  {Alley}(1988)}]{shih1988new}%
  \BibitemOpen
  \bibfield  {author} {\bibinfo {author} {\bibfnamefont {Y.~H.}\ \bibnamefont
  {Shih}}\ and\ \bibinfo {author} {\bibfnamefont {C.~O.}\ \bibnamefont
  {Alley}},\ }\bibfield  {title} {\bibinfo {title} {New type of
  einstein-podolsky-rosen-bohm experiment using pairs of light quanta produced
  by optical parametric down conversion},\ }\href@noop {} {\bibfield  {journal}
  {\bibinfo  {journal} {Phys. Rev. Lett.}\ }\textbf {\bibinfo {volume} {61}},\
  \bibinfo {pages} {2921} (\bibinfo {year} {1988})}\BibitemShut {NoStop}%
\bibitem [{\citenamefont {Rarity}\ and\ \citenamefont
  {Tapster}(1990)}]{rarity1990experimental}%
  \BibitemOpen
  \bibfield  {author} {\bibinfo {author} {\bibfnamefont {J.~G.}\ \bibnamefont
  {Rarity}}\ and\ \bibinfo {author} {\bibfnamefont {P.~R.}\ \bibnamefont
  {Tapster}},\ }\bibfield  {title} {\bibinfo {title} {Experimental violation of
  bell's inequality based on phase and momentum},\ }\href@noop {} {\bibfield
  {journal} {\bibinfo  {journal} {Phys. Rev. Lett.}\ }\textbf {\bibinfo
  {volume} {64}},\ \bibinfo {pages} {2495} (\bibinfo {year}
  {1990})}\BibitemShut {NoStop}%
\bibitem [{\citenamefont {Kwiat}\ \emph {et~al.}(1995)\citenamefont {Kwiat},
  \citenamefont {Mattle}, \citenamefont {Weinfurter}, \citenamefont
  {Zeilinger}, \citenamefont {Sergienko},\ and\ \citenamefont
  {Shih}}]{kwiat1995new}%
  \BibitemOpen
  \bibfield  {author} {\bibinfo {author} {\bibfnamefont {P.~G.}\ \bibnamefont
  {Kwiat}}, \bibinfo {author} {\bibfnamefont {K.}~\bibnamefont {Mattle}},
  \bibinfo {author} {\bibfnamefont {H.}~\bibnamefont {Weinfurter}}, \bibinfo
  {author} {\bibfnamefont {A.}~\bibnamefont {Zeilinger}}, \bibinfo {author}
  {\bibfnamefont {A.~V.}\ \bibnamefont {Sergienko}},\ and\ \bibinfo {author}
  {\bibfnamefont {Y.}~\bibnamefont {Shih}},\ }\bibfield  {title} {\bibinfo
  {title} {New high-intensity source of polarization-entangled photon pairs},\
  }\href@noop {} {\bibfield  {journal} {\bibinfo  {journal} {Phys. Rev. Lett.}\
  }\textbf {\bibinfo {volume} {75}},\ \bibinfo {pages} {4337} (\bibinfo {year}
  {1995})}\BibitemShut {NoStop}%
\bibitem [{\citenamefont {Weihs}\ \emph {et~al.}(1998)\citenamefont {Weihs},
  \citenamefont {Jennewein}, \citenamefont {Simon}, \citenamefont
  {Weinfurter},\ and\ \citenamefont {Zeilinger}}]{weihs1998violation}%
  \BibitemOpen
  \bibfield  {author} {\bibinfo {author} {\bibfnamefont {G.}~\bibnamefont
  {Weihs}}, \bibinfo {author} {\bibfnamefont {T.}~\bibnamefont {Jennewein}},
  \bibinfo {author} {\bibfnamefont {C.}~\bibnamefont {Simon}}, \bibinfo
  {author} {\bibfnamefont {H.}~\bibnamefont {Weinfurter}},\ and\ \bibinfo
  {author} {\bibfnamefont {A.}~\bibnamefont {Zeilinger}},\ }\bibfield  {title}
  {\bibinfo {title} {Violation of bell's inequality under strict einstein
  locality conditions},\ }\href@noop {} {\bibfield  {journal} {\bibinfo
  {journal} {Phys. Rev. Lett.}\ }\textbf {\bibinfo {volume} {81}},\ \bibinfo
  {pages} {5039} (\bibinfo {year} {1998})}\BibitemShut {NoStop}%
\bibitem [{\citenamefont {Rowe}\ \emph {et~al.}(2001)\citenamefont {Rowe},
  \citenamefont {Kielpinski}, \citenamefont {Meyer}, \citenamefont {Sackett},
  \citenamefont {Itano}, \citenamefont {Monroe},\ and\ \citenamefont
  {Wineland}}]{rowe2001experimental}%
  \BibitemOpen
  \bibfield  {author} {\bibinfo {author} {\bibfnamefont {M.~A.}\ \bibnamefont
  {Rowe}}, \bibinfo {author} {\bibfnamefont {D.}~\bibnamefont {Kielpinski}},
  \bibinfo {author} {\bibfnamefont {V.}~\bibnamefont {Meyer}}, \bibinfo
  {author} {\bibfnamefont {C.~A.}\ \bibnamefont {Sackett}}, \bibinfo {author}
  {\bibfnamefont {W.~M.}\ \bibnamefont {Itano}}, \bibinfo {author}
  {\bibfnamefont {C.}~\bibnamefont {Monroe}},\ and\ \bibinfo {author}
  {\bibfnamefont {D.~J.}\ \bibnamefont {Wineland}},\ }\bibfield  {title}
  {\bibinfo {title} {Experimental violation of a bell's inequality with
  efficient detection},\ }\href@noop {} {\bibfield  {journal} {\bibinfo
  {journal} {Nature}\ }\textbf {\bibinfo {volume} {409}},\ \bibinfo {pages}
  {791} (\bibinfo {year} {2001})}\BibitemShut {NoStop}%
\bibitem [{\citenamefont {Hensen}\ \emph {et~al.}(2015)\citenamefont {Hensen},
  \citenamefont {Bernien}, \citenamefont {Dr{\'e}au}, \citenamefont {Reiserer},
  \citenamefont {Kalb}, \citenamefont {Blok}, \citenamefont {Ruitenberg},
  \citenamefont {Vermeulen}, \citenamefont {Schouten}, \citenamefont
  {Abell{\'a}n} \emph {et~al.}}]{hensen2015loophole}%
  \BibitemOpen
  \bibfield  {author} {\bibinfo {author} {\bibfnamefont {B.}~\bibnamefont
  {Hensen}}, \bibinfo {author} {\bibfnamefont {H.}~\bibnamefont {Bernien}},
  \bibinfo {author} {\bibfnamefont {A.~E.}\ \bibnamefont {Dr{\'e}au}}, \bibinfo
  {author} {\bibfnamefont {A.}~\bibnamefont {Reiserer}}, \bibinfo {author}
  {\bibfnamefont {N.}~\bibnamefont {Kalb}}, \bibinfo {author} {\bibfnamefont
  {M.~S.}\ \bibnamefont {Blok}}, \bibinfo {author} {\bibfnamefont
  {J.}~\bibnamefont {Ruitenberg}}, \bibinfo {author} {\bibfnamefont {R.~F.}\
  \bibnamefont {Vermeulen}}, \bibinfo {author} {\bibfnamefont {R.~N.}\
  \bibnamefont {Schouten}}, \bibinfo {author} {\bibfnamefont {C.}~\bibnamefont
  {Abell{\'a}n}}, \emph {et~al.},\ }\bibfield  {title} {\bibinfo {title}
  {Loophole-free bell inequality violation using electron spins separated by
  1.3 kilometres},\ }\href@noop {} {\bibfield  {journal} {\bibinfo  {journal}
  {Nature}\ }\textbf {\bibinfo {volume} {526}},\ \bibinfo {pages} {682}
  (\bibinfo {year} {2015})}\BibitemShut {NoStop}%
\bibitem [{\citenamefont {Giustina}\ \emph {et~al.}(2015)\citenamefont
  {Giustina}, \citenamefont {Versteegh}, \citenamefont {Wengerowsky},
  \citenamefont {Handsteiner}, \citenamefont {Hochrainer}, \citenamefont
  {Phelan}, \citenamefont {Steinlechner}, \citenamefont {Kofler}, \citenamefont
  {Larsson}, \citenamefont {Abell\'an}, \citenamefont {Amaya}, \citenamefont
  {Pruneri}, \citenamefont {Mitchell}, \citenamefont {Beyer}, \citenamefont
  {Gerrits}, \citenamefont {Lita}, \citenamefont {Shalm}, \citenamefont {Nam},
  \citenamefont {Scheidl}, \citenamefont {Ursin}, \citenamefont {Wittmann},\
  and\ \citenamefont {Zeilinger}}]{PhysRevLett.115.250401}%
  \BibitemOpen
  \bibfield  {author} {\bibinfo {author} {\bibfnamefont {M.}~\bibnamefont
  {Giustina}}, \bibinfo {author} {\bibfnamefont {M.~A.~M.}\ \bibnamefont
  {Versteegh}}, \bibinfo {author} {\bibfnamefont {S.}~\bibnamefont
  {Wengerowsky}}, \bibinfo {author} {\bibfnamefont {J.}~\bibnamefont
  {Handsteiner}}, \bibinfo {author} {\bibfnamefont {A.}~\bibnamefont
  {Hochrainer}}, \bibinfo {author} {\bibfnamefont {K.}~\bibnamefont {Phelan}},
  \bibinfo {author} {\bibfnamefont {F.}~\bibnamefont {Steinlechner}}, \bibinfo
  {author} {\bibfnamefont {J.}~\bibnamefont {Kofler}}, \bibinfo {author}
  {\bibfnamefont {J.-A.}\ \bibnamefont {Larsson}}, \bibinfo {author}
  {\bibfnamefont {C.}~\bibnamefont {Abell\'an}}, \bibinfo {author}
  {\bibfnamefont {W.}~\bibnamefont {Amaya}}, \bibinfo {author} {\bibfnamefont
  {V.}~\bibnamefont {Pruneri}}, \bibinfo {author} {\bibfnamefont {M.~W.}\
  \bibnamefont {Mitchell}}, \bibinfo {author} {\bibfnamefont {J.}~\bibnamefont
  {Beyer}}, \bibinfo {author} {\bibfnamefont {T.}~\bibnamefont {Gerrits}},
  \bibinfo {author} {\bibfnamefont {A.~E.}\ \bibnamefont {Lita}}, \bibinfo
  {author} {\bibfnamefont {L.~K.}\ \bibnamefont {Shalm}}, \bibinfo {author}
  {\bibfnamefont {S.~W.}\ \bibnamefont {Nam}}, \bibinfo {author} {\bibfnamefont
  {T.}~\bibnamefont {Scheidl}}, \bibinfo {author} {\bibfnamefont
  {R.}~\bibnamefont {Ursin}}, \bibinfo {author} {\bibfnamefont
  {B.}~\bibnamefont {Wittmann}},\ and\ \bibinfo {author} {\bibfnamefont
  {A.}~\bibnamefont {Zeilinger}},\ }\bibfield  {title} {\bibinfo {title}
  {Significant-loophole-free test of bell's theorem with entangled photons},\
  }\href {https://doi.org/10.1103/PhysRevLett.115.250401} {\bibfield  {journal}
  {\bibinfo  {journal} {Phys. Rev. Lett.}\ }\textbf {\bibinfo {volume} {115}},\
  \bibinfo {pages} {250401} (\bibinfo {year} {2015})}\BibitemShut {NoStop}%
\bibitem [{\citenamefont {Shalm}\ \emph {et~al.}(2015)\citenamefont {Shalm},
  \citenamefont {Meyer-Scott}, \citenamefont {Christensen}, \citenamefont
  {Bierhorst}, \citenamefont {Wayne}, \citenamefont {Stevens}, \citenamefont
  {Gerrits}, \citenamefont {Glancy}, \citenamefont {Hamel}, \citenamefont
  {Allman} \emph {et~al.}}]{shalm2015strong}%
  \BibitemOpen
  \bibfield  {author} {\bibinfo {author} {\bibfnamefont {L.~K.}\ \bibnamefont
  {Shalm}}, \bibinfo {author} {\bibfnamefont {E.}~\bibnamefont {Meyer-Scott}},
  \bibinfo {author} {\bibfnamefont {B.~G.}\ \bibnamefont {Christensen}},
  \bibinfo {author} {\bibfnamefont {P.}~\bibnamefont {Bierhorst}}, \bibinfo
  {author} {\bibfnamefont {M.~A.}\ \bibnamefont {Wayne}}, \bibinfo {author}
  {\bibfnamefont {M.~J.}\ \bibnamefont {Stevens}}, \bibinfo {author}
  {\bibfnamefont {T.}~\bibnamefont {Gerrits}}, \bibinfo {author} {\bibfnamefont
  {S.}~\bibnamefont {Glancy}}, \bibinfo {author} {\bibfnamefont {D.~R.}\
  \bibnamefont {Hamel}}, \bibinfo {author} {\bibfnamefont {M.~S.}\ \bibnamefont
  {Allman}}, \emph {et~al.},\ }\bibfield  {title} {\bibinfo {title} {Strong
  loophole-free test of local realism},\ }\href@noop {} {\bibfield  {journal}
  {\bibinfo  {journal} {Phys. Rev. Lett.}\ }\textbf {\bibinfo {volume} {115}},\
  \bibinfo {pages} {250402} (\bibinfo {year} {2015})}\BibitemShut {NoStop}%
\bibitem [{\citenamefont {Vivoli}\ \emph {et~al.}(2016)\citenamefont {Vivoli},
  \citenamefont {Barnea}, \citenamefont {Galland},\ and\ \citenamefont
  {Sangouard}}]{vivoli2016proposal}%
  \BibitemOpen
  \bibfield  {author} {\bibinfo {author} {\bibfnamefont {V.~C.}\ \bibnamefont
  {Vivoli}}, \bibinfo {author} {\bibfnamefont {T.}~\bibnamefont {Barnea}},
  \bibinfo {author} {\bibfnamefont {C.}~\bibnamefont {Galland}},\ and\ \bibinfo
  {author} {\bibfnamefont {N.}~\bibnamefont {Sangouard}},\ }\bibfield  {title}
  {\bibinfo {title} {Proposal for an optomechanical bell test},\ }\href@noop {}
  {\bibfield  {journal} {\bibinfo  {journal} {Phys. Rev. Lett.}\ }\textbf
  {\bibinfo {volume} {116}},\ \bibinfo {pages} {070405} (\bibinfo {year}
  {2016})}\BibitemShut {NoStop}%
\bibitem [{\citenamefont {Hofer}\ \emph {et~al.}(2016)\citenamefont {Hofer},
  \citenamefont {Lehnert},\ and\ \citenamefont {Hammerer}}]{hofer2016proposal}%
  \BibitemOpen
  \bibfield  {author} {\bibinfo {author} {\bibfnamefont {S.~G.}\ \bibnamefont
  {Hofer}}, \bibinfo {author} {\bibfnamefont {K.~W.}\ \bibnamefont {Lehnert}},\
  and\ \bibinfo {author} {\bibfnamefont {K.}~\bibnamefont {Hammerer}},\
  }\bibfield  {title} {\bibinfo {title} {Proposal to test bell's inequality in
  electromechanics},\ }\href@noop {} {\bibfield  {journal} {\bibinfo  {journal}
  {Phys. Rev. Lett.}\ }\textbf {\bibinfo {volume} {116}},\ \bibinfo {pages}
  {070406} (\bibinfo {year} {2016})}\BibitemShut {NoStop}%
\bibitem [{\citenamefont {Marinkovi{\'c}}\ \emph {et~al.}(2018)\citenamefont
  {Marinkovi{\'c}}, \citenamefont {Wallucks}, \citenamefont {Riedinger},
  \citenamefont {Hong}, \citenamefont {Aspelmeyer},\ and\ \citenamefont
  {Gr{\"o}blacher}}]{marinkovic2018optomechanical}%
  \BibitemOpen
  \bibfield  {author} {\bibinfo {author} {\bibfnamefont {I.}~\bibnamefont
  {Marinkovi{\'c}}}, \bibinfo {author} {\bibfnamefont {A.}~\bibnamefont
  {Wallucks}}, \bibinfo {author} {\bibfnamefont {R.}~\bibnamefont {Riedinger}},
  \bibinfo {author} {\bibfnamefont {S.}~\bibnamefont {Hong}}, \bibinfo {author}
  {\bibfnamefont {M.}~\bibnamefont {Aspelmeyer}},\ and\ \bibinfo {author}
  {\bibfnamefont {S.}~\bibnamefont {Gr{\"o}blacher}},\ }\bibfield  {title}
  {\bibinfo {title} {Optomechanical bell test},\ }\href@noop {} {\bibfield
  {journal} {\bibinfo  {journal} {Phys. Rev. Lett.}\ }\textbf {\bibinfo
  {volume} {121}},\ \bibinfo {pages} {220404} (\bibinfo {year}
  {2018})}\BibitemShut {NoStop}%
\bibitem [{\citenamefont {Paris}(1996)}]{paris1996displacement}%
  \BibitemOpen
  \bibfield  {author} {\bibinfo {author} {\bibfnamefont {M.~G.~A.}\
  \bibnamefont {Paris}},\ }\bibfield  {title} {\bibinfo {title} {Displacement
  operator by beam splitter},\ }\href@noop {} {\bibfield  {journal} {\bibinfo
  {journal} {Phys. Lett. A}\ }\textbf {\bibinfo {volume} {217}},\ \bibinfo
  {pages} {78} (\bibinfo {year} {1996})}\BibitemShut {NoStop}%
\bibitem [{\citenamefont {Kuzmich}\ \emph {et~al.}(2000)\citenamefont
  {Kuzmich}, \citenamefont {Walmsley},\ and\ \citenamefont
  {Mandel}}]{kuzmich2000violation}%
  \BibitemOpen
  \bibfield  {author} {\bibinfo {author} {\bibfnamefont {A.}~\bibnamefont
  {Kuzmich}}, \bibinfo {author} {\bibfnamefont {I.~A.}\ \bibnamefont
  {Walmsley}},\ and\ \bibinfo {author} {\bibfnamefont {L.}~\bibnamefont
  {Mandel}},\ }\bibfield  {title} {\bibinfo {title} {Violation of bell's
  inequality by a generalized einstein-podolsky-rosen state using homodyne
  detection},\ }\href@noop {} {\bibfield  {journal} {\bibinfo  {journal} {Phys.
  Rev. Lett.}\ }\textbf {\bibinfo {volume} {85}},\ \bibinfo {pages} {1349}
  (\bibinfo {year} {2000})}\BibitemShut {NoStop}%
\bibitem [{\citenamefont {Wiseman}\ and\ \citenamefont
  {Milburn}(1994)}]{PhysRevA.49.4110}%
  \BibitemOpen
  \bibfield  {author} {\bibinfo {author} {\bibfnamefont {H.~M.}\ \bibnamefont
  {Wiseman}}\ and\ \bibinfo {author} {\bibfnamefont {G.~J.}\ \bibnamefont
  {Milburn}},\ }\bibfield  {title} {\bibinfo {title} {All-optical versus
  electro-optical quantum-limited feedback},\ }\href
  {https://doi.org/10.1103/PhysRevA.49.4110} {\bibfield  {journal} {\bibinfo
  {journal} {Phys. Rev. A}\ }\textbf {\bibinfo {volume} {49}},\ \bibinfo
  {pages} {4110} (\bibinfo {year} {1994})}\BibitemShut {NoStop}%
\bibitem [{\citenamefont {Hofer}\ \emph {et~al.}(2011)\citenamefont {Hofer},
  \citenamefont {Wieczorek}, \citenamefont {Aspelmeyer},\ and\ \citenamefont
  {Hammerer}}]{hofer2011quantum}%
  \BibitemOpen
  \bibfield  {author} {\bibinfo {author} {\bibfnamefont {S.~G.}\ \bibnamefont
  {Hofer}}, \bibinfo {author} {\bibfnamefont {W.}~\bibnamefont {Wieczorek}},
  \bibinfo {author} {\bibfnamefont {M.}~\bibnamefont {Aspelmeyer}},\ and\
  \bibinfo {author} {\bibfnamefont {K.}~\bibnamefont {Hammerer}},\ }\bibfield
  {title} {\bibinfo {title} {Quantum entanglement and teleportation in pulsed
  cavity optomechanics},\ }\href@noop {} {\bibfield  {journal} {\bibinfo
  {journal} {Phys. Rev. A}\ }\textbf {\bibinfo {volume} {84}},\ \bibinfo
  {pages} {052327} (\bibinfo {year} {2011})}\BibitemShut {NoStop}%
\bibitem [{\citenamefont {Galland}\ \emph {et~al.}(2014)\citenamefont
  {Galland}, \citenamefont {Sangouard}, \citenamefont {Piro}, \citenamefont
  {Gisin},\ and\ \citenamefont {Kippenberg}}]{galland2014heralded}%
  \BibitemOpen
  \bibfield  {author} {\bibinfo {author} {\bibfnamefont {C.}~\bibnamefont
  {Galland}}, \bibinfo {author} {\bibfnamefont {N.}~\bibnamefont {Sangouard}},
  \bibinfo {author} {\bibfnamefont {N.}~\bibnamefont {Piro}}, \bibinfo {author}
  {\bibfnamefont {N.}~\bibnamefont {Gisin}},\ and\ \bibinfo {author}
  {\bibfnamefont {T.~J.}\ \bibnamefont {Kippenberg}},\ }\bibfield  {title}
  {\bibinfo {title} {Heralded single-phonon preparation, storage, and readout
  in cavity optomechanics},\ }\href@noop {} {\bibfield  {journal} {\bibinfo
  {journal} {Phys. Rev. Lett.}\ }\textbf {\bibinfo {volume} {112}},\ \bibinfo
  {pages} {143602} (\bibinfo {year} {2014})}\BibitemShut {NoStop}%
\bibitem [{\citenamefont {Tan}\ \emph {et~al.}(1991)\citenamefont {Tan},
  \citenamefont {Walls},\ and\ \citenamefont {Collett}}]{tan1991nonlocality}%
  \BibitemOpen
  \bibfield  {author} {\bibinfo {author} {\bibfnamefont {S.~M.}\ \bibnamefont
  {Tan}}, \bibinfo {author} {\bibfnamefont {D.~F.}\ \bibnamefont {Walls}},\
  and\ \bibinfo {author} {\bibfnamefont {M.~J.}\ \bibnamefont {Collett}},\
  }\bibfield  {title} {\bibinfo {title} {Nonlocality of a single photon},\
  }\href@noop {} {\bibfield  {journal} {\bibinfo  {journal} {Phys. Rev. Lett.}\
  }\textbf {\bibinfo {volume} {66}},\ \bibinfo {pages} {252} (\bibinfo {year}
  {1991})}\BibitemShut {NoStop}%
\bibitem [{\citenamefont {Banaszek}\ and\ \citenamefont
  {W{\'o}dkiewicz}(1999)}]{banaszek1999testing}%
  \BibitemOpen
  \bibfield  {author} {\bibinfo {author} {\bibfnamefont {K.}~\bibnamefont
  {Banaszek}}\ and\ \bibinfo {author} {\bibfnamefont {K.}~\bibnamefont
  {W{\'o}dkiewicz}},\ }\bibfield  {title} {\bibinfo {title} {Testing quantum
  nonlocality in phase space},\ }\href@noop {} {\bibfield  {journal} {\bibinfo
  {journal} {Phys. Rev. Lett.}\ }\textbf {\bibinfo {volume} {82}},\ \bibinfo
  {pages} {2009} (\bibinfo {year} {1999})}\BibitemShut {NoStop}%
\bibitem [{\citenamefont {Hardy}(1994)}]{hardy1994nonlocality}%
  \BibitemOpen
  \bibfield  {author} {\bibinfo {author} {\bibfnamefont {L.}~\bibnamefont
  {Hardy}},\ }\bibfield  {title} {\bibinfo {title} {Nonlocality of a single
  photon revisited},\ }\href@noop {} {\bibfield  {journal} {\bibinfo  {journal}
  {Phys. Rev. Lett.}\ }\textbf {\bibinfo {volume} {73}},\ \bibinfo {pages}
  {2279} (\bibinfo {year} {1994})}\BibitemShut {NoStop}%
\bibitem [{\citenamefont {Hessmo}\ \emph {et~al.}(2004)\citenamefont {Hessmo},
  \citenamefont {Usachev}, \citenamefont {Heydari},\ and\ \citenamefont
  {Bj{\"o}rk}}]{hessmo2004experimental}%
  \BibitemOpen
  \bibfield  {author} {\bibinfo {author} {\bibfnamefont {B.}~\bibnamefont
  {Hessmo}}, \bibinfo {author} {\bibfnamefont {P.}~\bibnamefont {Usachev}},
  \bibinfo {author} {\bibfnamefont {H.}~\bibnamefont {Heydari}},\ and\ \bibinfo
  {author} {\bibfnamefont {G.}~\bibnamefont {Bj{\"o}rk}},\ }\bibfield  {title}
  {\bibinfo {title} {Experimental demonstration of single photon nonlocality},\
  }\href@noop {} {\bibfield  {journal} {\bibinfo  {journal} {Phys. Rev. Lett.}\
  }\textbf {\bibinfo {volume} {92}},\ \bibinfo {pages} {180401} (\bibinfo
  {year} {2004})}\BibitemShut {NoStop}%
\bibitem [{\citenamefont {Lee}\ \emph {et~al.}(2009)\citenamefont {Lee},
  \citenamefont {Jeong},\ and\ \citenamefont {Jaksch}}]{lee2009testing}%
  \BibitemOpen
  \bibfield  {author} {\bibinfo {author} {\bibfnamefont {S.-W.}\ \bibnamefont
  {Lee}}, \bibinfo {author} {\bibfnamefont {H.}~\bibnamefont {Jeong}},\ and\
  \bibinfo {author} {\bibfnamefont {D.}~\bibnamefont {Jaksch}},\ }\bibfield
  {title} {\bibinfo {title} {Testing quantum nonlocality by generalized
  quasiprobability functions},\ }\href@noop {} {\bibfield  {journal} {\bibinfo
  {journal} {Phys. Rev. A}\ }\textbf {\bibinfo {volume} {80}},\ \bibinfo
  {pages} {022104} (\bibinfo {year} {2009})}\BibitemShut {NoStop}%
\bibitem [{\citenamefont {Brask}\ and\ \citenamefont
  {Chaves}(2012)}]{brask2012robust}%
  \BibitemOpen
  \bibfield  {author} {\bibinfo {author} {\bibfnamefont {J.~B.}\ \bibnamefont
  {Brask}}\ and\ \bibinfo {author} {\bibfnamefont {R.}~\bibnamefont {Chaves}},\
  }\bibfield  {title} {\bibinfo {title} {Robust nonlocality tests with
  displacement-based measurements},\ }\href@noop {} {\bibfield  {journal}
  {\bibinfo  {journal} {Phys. Rev. A}\ }\textbf {\bibinfo {volume} {86}},\
  \bibinfo {pages} {010103(R)} (\bibinfo {year} {2012})}\BibitemShut {NoStop}%
\bibitem [{\citenamefont {Li}\ and\ \citenamefont
  {Zhu}(2017)}]{li2017einstein}%
  \BibitemOpen
  \bibfield  {author} {\bibinfo {author} {\bibfnamefont {J.}~\bibnamefont
  {Li}}\ and\ \bibinfo {author} {\bibfnamefont {S.-Y.}\ \bibnamefont {Zhu}},\
  }\bibfield  {title} {\bibinfo {title} {Einstein-podolsky-rosen steering and
  bell nonlocality of two macroscopic mechanical oscillators in optomechanical
  systems},\ }\href@noop {} {\bibfield  {journal} {\bibinfo  {journal} {Phys.
  Rev. A}\ }\textbf {\bibinfo {volume} {96}},\ \bibinfo {pages} {062115}
  (\bibinfo {year} {2017})}\BibitemShut {NoStop}%
\end{thebibliography}%

\end{document}